\documentclass[3p,twocolumn]{elsarticle} %
\usepackage[utf8]{inputenc}
\usepackage{hyperref}
\usepackage[intlimits]{amsmath}
\usepackage{amsfonts}
\usepackage{bm}
\usepackage{graphicx}
\usepackage{physics}

\usepackage{subcaption}
\usepackage{listings}
\usepackage{float}

\DeclareMathOperator{\naive}{naive}
\DeclareMathOperator{\fitted}{fitted}
\DeclareMathOperator{\Bee}{Bee}
\DeclareMathOperator{\invariant}{invariant}

\begin{document}

\title{Robust test statistics for data sets with missing correlation information}
\journal{PRD}

\author{Lukas Koch}
\address{University of Oxford}
\ead{lukas.koch@physics.ox.ac.uk}

\begin{abstract}
Not all experiments publish their results with a description of the correlations between the data points.
This makes it difficult to do hypothesis tests or model fits with that data,
since just assuming no correlation can lead to an over- or underestimation of the resulting uncertainties.
This work presents robust test statistics that can be used with data sets with missing correlation information.
They are exact in the case of no correlation and either guaranteed to be conservative -- i.e. the uncertainty is never underestimated -- in the presence of correlations,
or they are also exact in the degenerate case of perfect correlation between the data points.
\end{abstract}

\maketitle

\section{Introduction}
\label{sec:introduction}

Some data sets are published without a full covariance matrix, describing the correlations between the data points of the result.
The implied assumption in these data sets is that the correlation in the uncertainties is 0, i.e. the data is uncorrelated.
This is not always the case though\footnote{If, e.g., the data points vary smoothly within their error bands or when they are supposed to describe a ``shape-only'' uncertainty, it is clear that there is a correlation there.},
and users of the data are put in the unenviable situation of having to use correlated results,
without knowing what the correlations actually are.
Usually one would use the fully correlated Mahalanobis distance\cite{Mahalanobis1936,Wilks2019a} or its square, $D^2 = \bm{\Delta}^T S^{-1} \bm{\Delta}$, to judge how well a certain model fits the data\footnote{In the particle physics community, this is often simply called ``the chi-square''. To avoid confusion with other chi-square distributed test statistics, it is useful to use its proper name, though.}.
Here $\bm{\Delta}$ is the difference between the data and the model prediction,
and $S$ is the covariance matrix describing the uncertainty of the result.
Just ignoring the correlations and applying a ``naive'', uncorrelated Mahalanobis distance to compare a model to the data can lead to plainly wrong results.

\begin{figure*}
    \centering
    \includegraphics[width=0.49\textwidth]{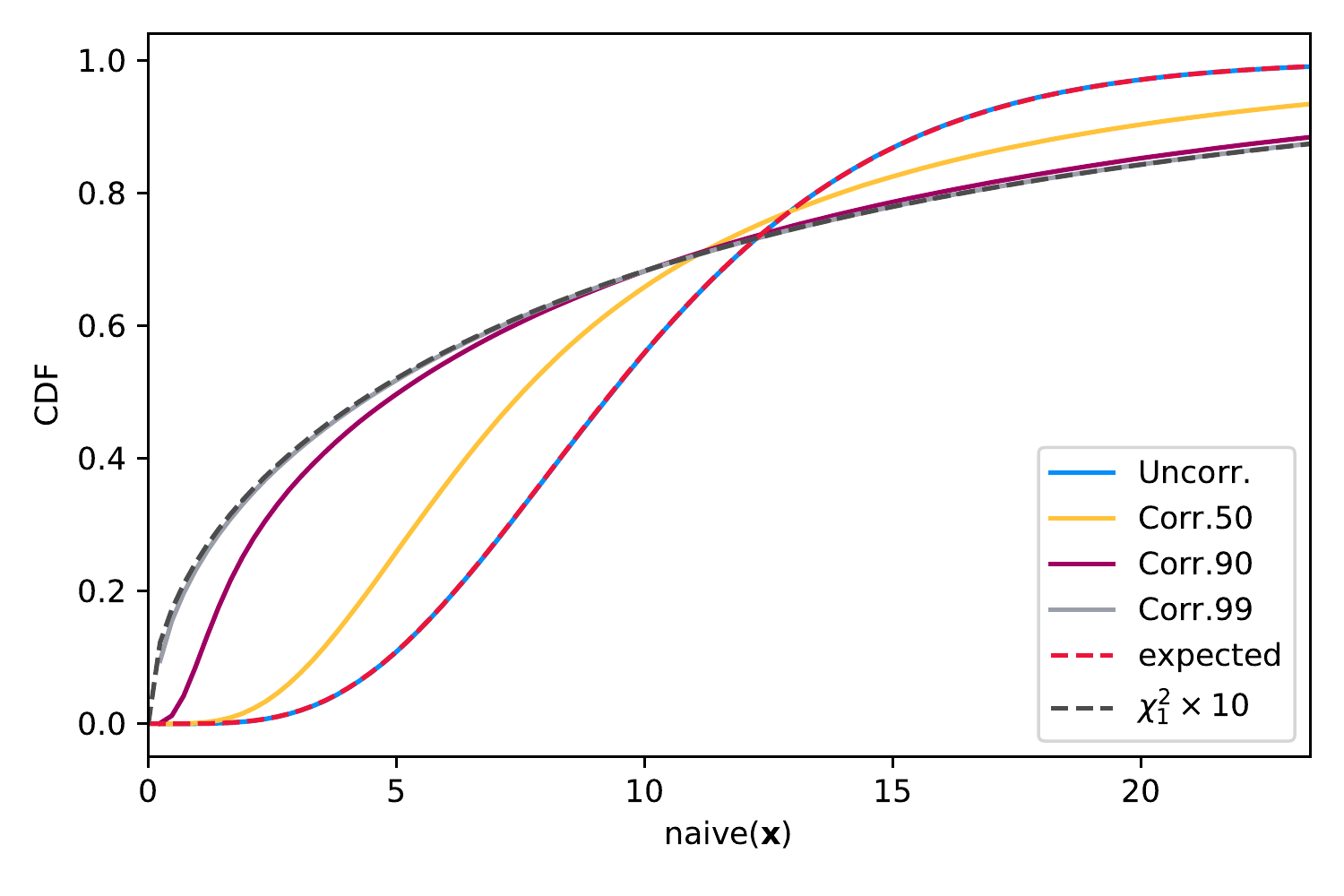}
    \includegraphics[width=0.49\textwidth]{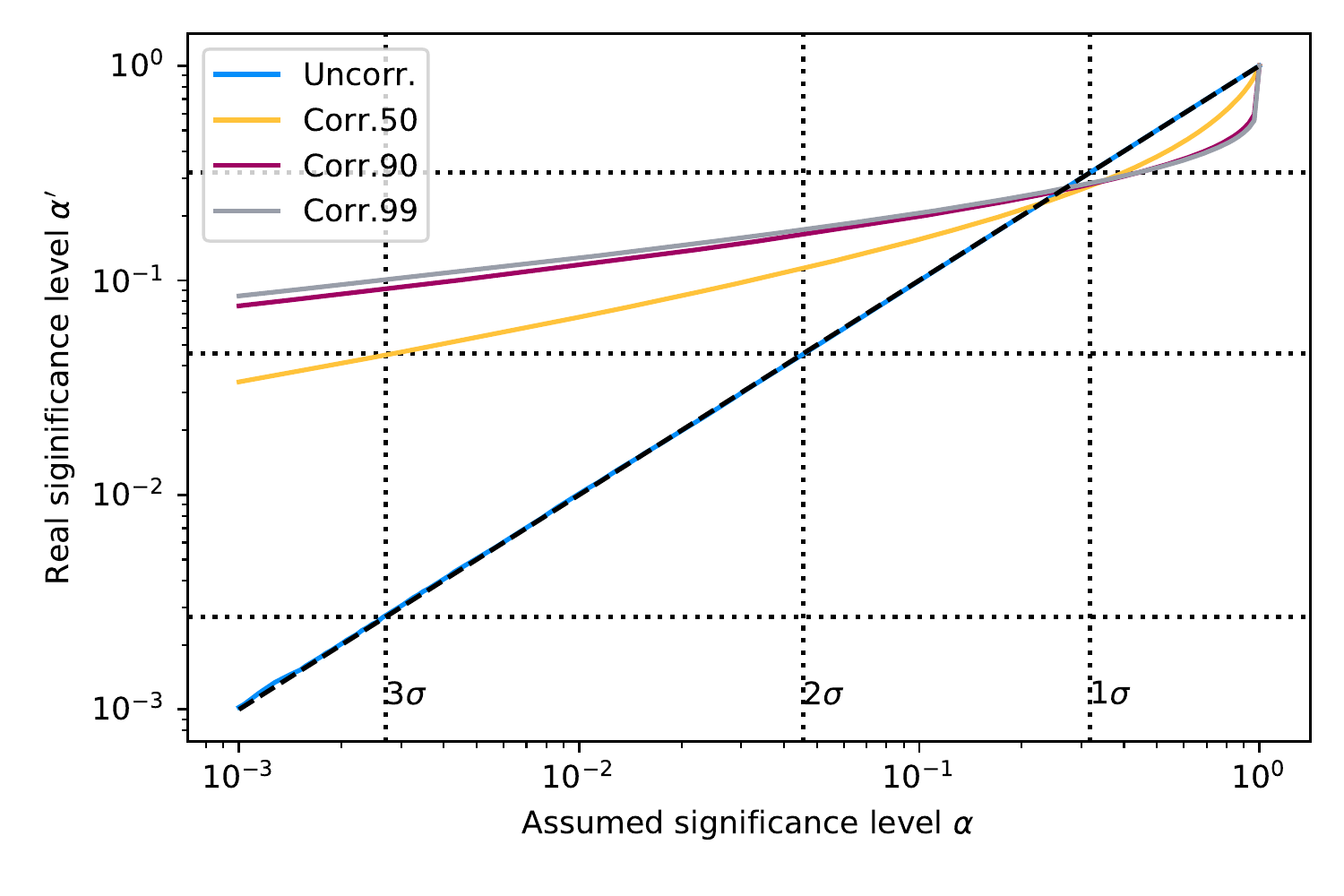}
    \caption{\label{fig:naive}%
        CDFs (left) for the ``naive'' test statistic for different levels of correlations in the data.
        When using the uncorrelated CDF to calculate the assumed significance level (or p-value) of a value of the statistic, the actual level will differ from the assumption depending on the correlations (right).
        As the correlation increases, the distribution of the naive test statistic approaches that of a $\chi^2_1$ distributed variable which is multiplied by the number of bins (in this case 10).
    }
\end{figure*}

Consider multivariate normal distributed data with ten dimensions.
In fact, unless otherwise stated, let us assume that the examples in this paper are all multivariate normal distributed and we know the correct diagonal elements of the respective covariance matrices.
The only problem we will address here is the missing of information regarding the correlations between the variables.
The ``naive'' test statistic would consist of just summing up the squared z-scores, i.e. the residuals normalised by the uncertainty:
\begin{equation}
    \naive(\bm{x}\,|\,\bm{\mu},\bm{s}) = \sum_i \frac{(x_i - \mu_i)^2}{S_{ii}}\text{,}
\end{equation}
where $\bm{x}$ is the data result, $\bm{\mu}$ is a prediction of the expectation value from some model, and $S_{ii} = s^2_i$ is the variance of the data points.
This test statistic will be chi-square distributed if there are no correlations present in the data (see any introductory statistics text book, e.g. \cite[ch. 6]{Rice2006}),
but using it in the presence of correlations leads to under- or overestimation of uncertainty,
depending on the correlation and the actual value of the statistic.
\autoref{fig:naive} shows this for 10-dimensional toy data sets thrown with different levels of correlation.
The diagonal terms of the covariance matrix are kept constant at 1, while the off-diagonals are set to the values 0, 0.5, 0.9, and 0.99, in the different sets.

The left plot shows the cumulative probability density functions (CDFs) of the expected distribution of the test statistic in the absence of correlations, as well as the actual CDFs of the different toy data sets.
The different distributions affect which significance level (or p-value) a certain value of the test statistic corresponds to.
The right plot shows how an assumed significance level (as calculated with the expected CDF) translates to the actual significance level (as calculated from the actual CDFs).
To put it another way: The x-axis shows how often one would like to make a Type-I error (rejecting a true hypothesis), while the y-axis shows how often one actually makes a Type-I error, given the different levels of correlation in the data.
If the actual significance level is larger than assumed,
one rejects a true hypothesis more often than intended.
In terms of error bars or confidence regions, this means that the size of the uncertainties is effectively underestimated.

This behaviour is clearly undesirable.
If the data is (suspected to be) strongly correlated,
it would be better to use a different test statistic that is able to perform consistently under different levels of correlation in the data.
For such a test statistic, the following properties would be desirable:
\begin{enumerate}
    \item Exact in the case of no correlations. \label{itm:nocorr}
    \item Guaranteed to be conservative when not exact.
    \item Low deviations from exactness when not exact.
    \item Exact in the case of 100\% correlation.
    \item Exact at every possible level of correlation.
\end{enumerate}
Some of these properties are contained in one another.
The naive use of the uncorrelated Mahalanobis distance has property~\ref{itm:nocorr} but none of the others.
The following sections will describe some test statistics that have more of these properties.
After that, we will compare them in \autoref{sec:comparison}, and apply them to some real experimental results from neutrino scattering experiments in \autoref{sec:application}.

\section{Fitting the covariance to the data}

The problem of estimating both the mean and covariance of multivariate normal distributed data has been extensively discussed in statistical literature (see e.g. \cite{TsukumaKubokawa2020} and references therein).
This includes work on estimators for the covariance when the number of observations is smaller than the number of dimensions of the data space\cite{Tsukuma2016}.
Unfortunately, the problem addressed here is somewhat unique, since we only have access to a single observation from the distribution we would like to estimate.
This observation is the published result.
We cannot assume that the models we try to test are drawn from the same distribution, since this is exactly the hypothesis we want to test.
Another difference to the widely discussed case -- this time in our favour -- is that we can assume to know the diagonal elements of the covariance matrix as well as the mean values of the distribution.
Also, we are not interested in the actual values of the full covariance matrix, as long as we can construct a test statistic that performs well without knowing these values.

The first considered test statistic thus arises from treating the off-diagonal elements of the covariance matrix as nuisance parameters of the statistical model.
For any given predicted mean value in the N-dimensional data space $\bm{\mu}$ and a given sample (i.e. the data) $\bm{x}$,
it is possible to choose the off-diagonal elements of $S$ in a way to minimise the resulting squared Mahalanobis distance $D^2 = (\bm{x} - \bm{\mu)^T}S^{-1}(\bm{x} - \bm{\mu})$.
This is different from maximising the likelihood of the data,
since the probability density of a multivariate normal distribution also depends on the determinant of the covariance matrix:
\begin{equation}
    L = (2\pi )^{-{\frac {N}{2}}}\det(S)^{-{\frac {1}{2}}}\,e^{-{\frac {1}{2}}(\mathbf {x} -{\boldsymbol {\mu }})^{\!{\mathsf {T}}}{S^{-1}(\mathbf {x} -{\boldsymbol {\mu }})}}\text{.}
\end{equation}
A minimal Mahalanobis distance is only equivalent to a maximal likelihood if the determinant of the covariance matrix is constant.
This is not the case here.
Furthermore, for $N \ge 3$ the supremum of the likelihood in a maximisation over the covariance elements is always $+\infty$,
rendering it useless as a test statistic.
This will be shown below.

Since the Mahalanobis distance is invariant under a linear transformation of the variables\cite{Wilks2019a,Bhattacharya2016a},
we can simplify the minimisation by transforming the variable space to make the last variable $x_N$ independent of the others:
\begin{align}
    \bm{y} &= \mqty*(\bm{1} & - \frac{\bm{S}_N}{S_{NN}} \\ 0 & 1) \bm{x} \\
    &= \mqty*(1 & 0 & \dots & 0 & - \frac{S_{1N}}{S_{NN}} \\
             0 & 1 & \dots & 0 & - \frac{S_{2N}}{S_{NN}} \\
             \vdots & \vdots & \ddots & \vdots &\vdots \\
             0 & 0 & \dots & 1 & - \frac{S_{(N-1)N}}{S_{NN}}\\
             0 & 0 & \dots & 0 & 1 )
             \bm{x} \text{.}
\end{align}
Here $S_{NN}$ is the variance of the Nth variable, and $\bm{S}_N$ is the vector of the $N-1$ covariances between the Nth and the other variables:
\begin{equation}
    \bm{S}_N = \mqty*(S_{1N} \\ \vdots \\ S_{(N-1)N})\text{.}
\end{equation}
The covariance and expectation values for $\bm{y}$ are then:
\begin{align}
    S^y &= (\grad \bm{y}^T)^T S (\grad \bm{y}^T) \\
        &= \mqty*(\bm{1} & - \frac{\bm{S}_N}{S_{NN}} \\ 0 & 1)
            \mqty*(S^{/N} & \bm{S}_N \\ \bm{S}^T_N & S_{NN})
            \mqty*(\bm{1} & 0 \\ - \frac{\bm{S}^T_N}{S_{NN}} & 1) \\
        &= \mqty*(S^{/N} - \frac{\bm{S}_N\bm{S}_N^T}{S_{NN}} & 0 \\ \bm{S}^T_N & S_{NN})
            \mqty*(\bm{1} & 0 \\ - \frac{\bm{S}^T_N}{S_{NN}} & 1) \\    
        &= \mqty*(S^{/N} - \frac{\bm{S}_N\bm{S}_N^T}{S_{NN}} & 0 \\ 0 & S_{NN})\text{,} \\
    \bm{\mu}^y &= \mqty*(\bm{1} & - \frac{\bm{S}_N}{S_{NN}} \\ 0 & 1) \bm{\mu}\text{,}
\end{align}
where $S^{/N}$ is the original covariance matrix for the remaining $N-1$ variables.

The contribution of $y_N = x_N$ to the total Mahalanobis distance $D^2 = (\bm{y} - \bm{\mu}^y)^T(S^y)^{-1}(\bm{y} - \bm{\mu}^y)$ is fixed, since it only depends on $S_{NN}$, which has a given constant value.
The contribution of the remaining variables $\bm{y}^{/N}$ could be minimised to 0 by choosing the off-diagonal elements of the covariance such that their expectation value $\bm{\mu}^{y/N}$ is equal to the actual value:
\begin{gather}
    \bm{\mu}^{y/N} \overset{!}{=} \bm{y}^{/N} \\
    \bm{\mu}^{/N} - \frac{\bm{S}_N}{S_{NN}} \mu_N = \bm{x}^{/N} - \frac{\bm{S}_N}{S_{NN}} x_N
\end{gather}
\begin{align}
    \bm{S}_N &= S_{NN} \frac{\bm{x}^{/N} - \bm{\mu}^{/N}}{x_N - \mu_N} \\
    &= \mqty*(\frac{\Delta_i}{\sqrt{S_{ii}}} \frac{\sqrt{S_{NN}}}{\Delta_N} \sqrt{S_{ii}S_{NN}} \\ \vdots \\ \frac{\Delta_{N-1}}{\sqrt{S_{(N-1)(N-1)}}} \frac{\sqrt{S_{NN}}}{\Delta_N} \sqrt{S_{(N-1)(N-1)}S_{NN}})
\end{align}
Here $\Delta_i/\sqrt{S_{ii}} = (x_i - \mu_i)/\sqrt{S_{ii}}$ is the (positive or negative) z-score of the $i$th variable, and $\sqrt{S_{ii}S_{NN}}$ is the maximum allowed absolute value of the covariance between the $i$th and $N$th variable.
The latter arises from the fact that the correlation coefficients $S_{ij}/\sqrt{S_{ii}S_{NN}}$ must be within $[-1, 1]$.
The vector of covariances $\bm{S}_N$ is thus the ratio of the $N-1$ z-scores over the Nth z-score, multiplied by the maximum allowed value for each covariance.
This is a valid choice of covariances when the Nth variable has the largest absolute z-score,
meaning that all z-score ratios are within $[-1, 1]$.

\begin{figure*}
    \centering
    \includegraphics[width=\textwidth]{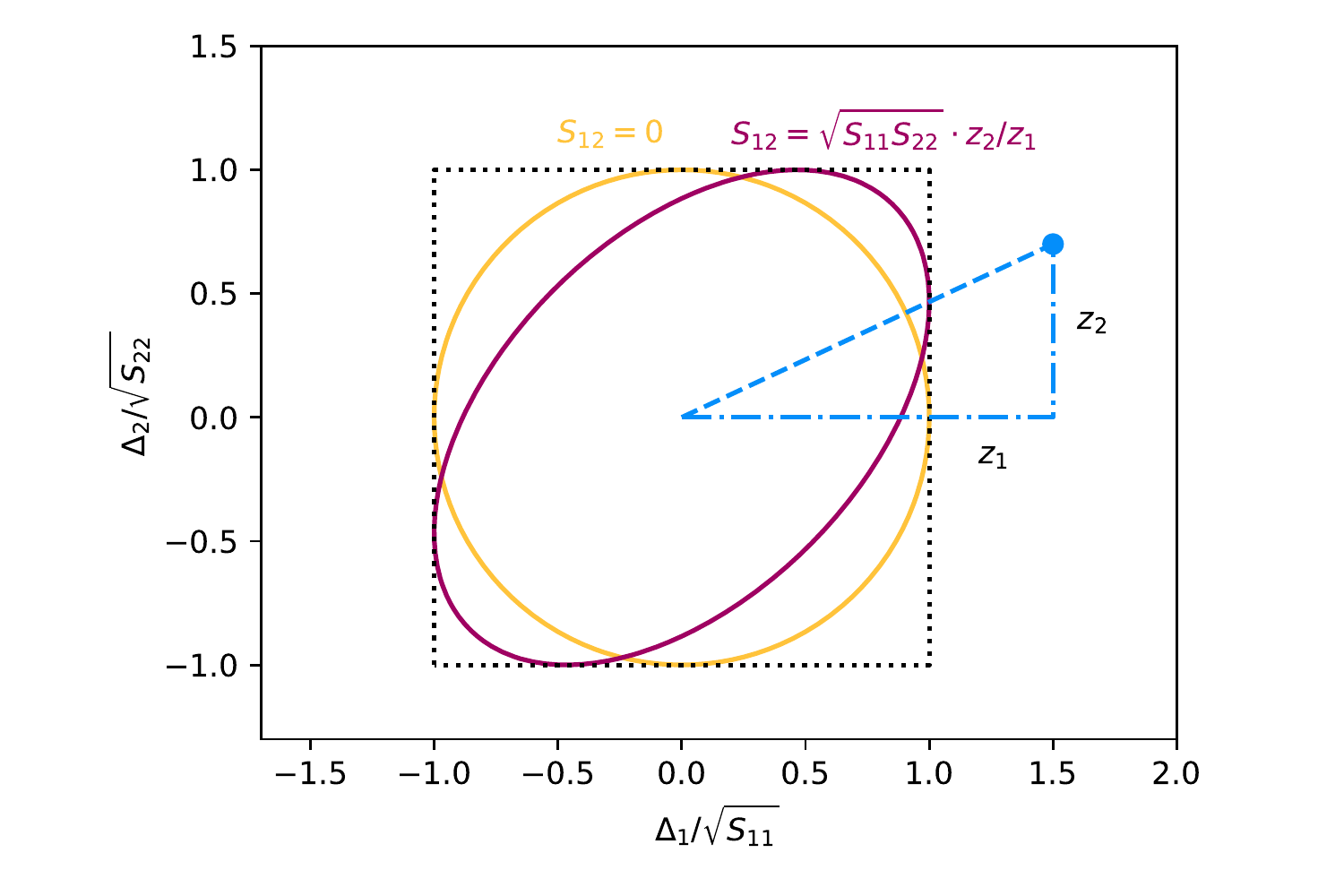}
    \caption{\label{fig:min-dist}%
    Minimum achievable Mahalanobis distance for two dimensions.
    The minimum achievable Mahalanobis distance when varying the off-diagonal covariance element is equal to the largest absolute z-score of the single variables.
    The surface where the Mahalanobis distance is equal to 1 is an ellipse contained within the square with its edges at $\Delta_i / \sqrt{S_{ii}} = \pm 1$.
    Varying the off-diagonal element of the covariance matrix does not rotate the principal axes of the ellipse, but it changes where it touches the edges of the square.
    When chosen correctly, the ellipse touches the edge of the box at the point where the data is projected onto it.
    Because of the linearity of the Mahalanobis distance, this means that the total distance of the data point is then simply the largest z-score.
    This two-dimensional minimisation can be done for all marginal  projections of pairs of variables in N-dimensional problems.
    To achieve the minimal total Mahalanobis distance, only the pairs involving the overall largest absolute z-score need to be specified like this,
    but applying the scheme to all pairs ensures a valid covariance matrix.
    }
\end{figure*}

We can always reorder the variables such that the $N$th is the one with the largest absolute z-score.
Thus, by eliminating the contribution of the other variables as shown above, the minimal achievable Mahalanobis distance under variation of the off-diagonal elements of the covariance matrix is equal to the largest absolute z-score of the single variables.
This behaviour is illustrated in \autoref{fig:min-dist} for two dimensions.
Note that we only need $N - 1$ of the $N(N-1)/2$ covariance parameters to ensure the value of the Mahalanobis distance.
The remaining elements can be chosen freely as long as they result in a valid covariance matrix.
In fact, they could be chosen in a way to make the determinant of $S$ arbitrarily small,
leading to the infinite supremum of the likelihood maximisation over the covariance elements for $N \ge 3$.

Since the minimum achievable Mahalanobis distance is always equal to the maximum absolute z-score among the variables,
no actual fitting or optimisation needs to be done for this test statistic.
Let us call $b$ the largest absolute z-score,
and we can define the ``fitted'' test statistic as:
\begin{equation}
    \fitted(\bm{\Delta}\,|\,\bm{s}) = b^2 =  \max_i\qty(\frac{\Delta_i^2}{S_{ii}} )\text{.}
\end{equation}

It is straight-forward to derive the expected distribution of this test statistic in the case of no correlations.
The CDF of $b$, $F_b(b')$, is just the probability of the absolute values of \emph{all} z-scores being smaller than or equal to $b'$:
\begin{gather}
    F_b(b') = P(b \le b') \nonumber = \\
    \int_{-b'}^{+b'} \dots \int_{-b'}^{+b'} f(z_1,\dots,z_N) \dd{z_1} \dots \dd{z_N} \text{,}
\end{gather}
where $f(\bm{z})$ denotes the probability density function (PDF) of the potentially negative z-scores.
With uncorrelated, standard normal distributed z-scores this evaluates to
\begin{equation}
    F_b(b') = \erf^N\qty(\frac{b'}{\sqrt{2}}) \text{.}
\end{equation}
With this we can write down the CDF of $b^2$ as
\begin{equation}
    F_{b^2}(y) = F_b(\sqrt{y}) \text{,}
\end{equation}
which defines the distribution of the fitted test statistic.
We will call the distribution the ``Bee-square'' distribution (as a nod to the chi-square distribution) and we have
\begin{equation}
    \fitted(\bm{\Delta}\,|\,\bm{s}) \sim \Bee^2_N \text{,}
\end{equation}
for uncorrelated normal distributed $\Delta_i$.
A Python implementation of the distribution can be found in Listing~\ref{lst:bee} in the appendix.

\autoref{fig:fitted} shows how this test statistic fares for different levels of correlation in the toy data.
For no correlations, the distribution follows the expectation.
With increasing correlations, the distribution deviates more and more,
approaching a chi-square distribution with one degree of freedom.\footnote{As the data gets more and more correlated, the z-scores will approach being equal in all cases, and the maximum z-score will be distributed like a single standard normal distributed variable.}
Compared to the naive approach, we can see that the deviation from an exact statistic has been decreased for a wide range of significance levels.
But more importantly, the fitted test statistic is conservative for all significance levels and all correlation strengths.
The real significance level of a result is always equal to or lower than the assumed significance that was evaluated using the expected Bee-square distribution.\footnote{I.e. the probability of a result at least as extreme as the observed is actually lower than what the assumed distribution suggest; the uncertainty is overestimated.}
A proof of this for the two-dimensional case can be found in \ref{sec:fitted-proof}.

\begin{figure*}
    \centering
    \includegraphics[width=0.49\textwidth]{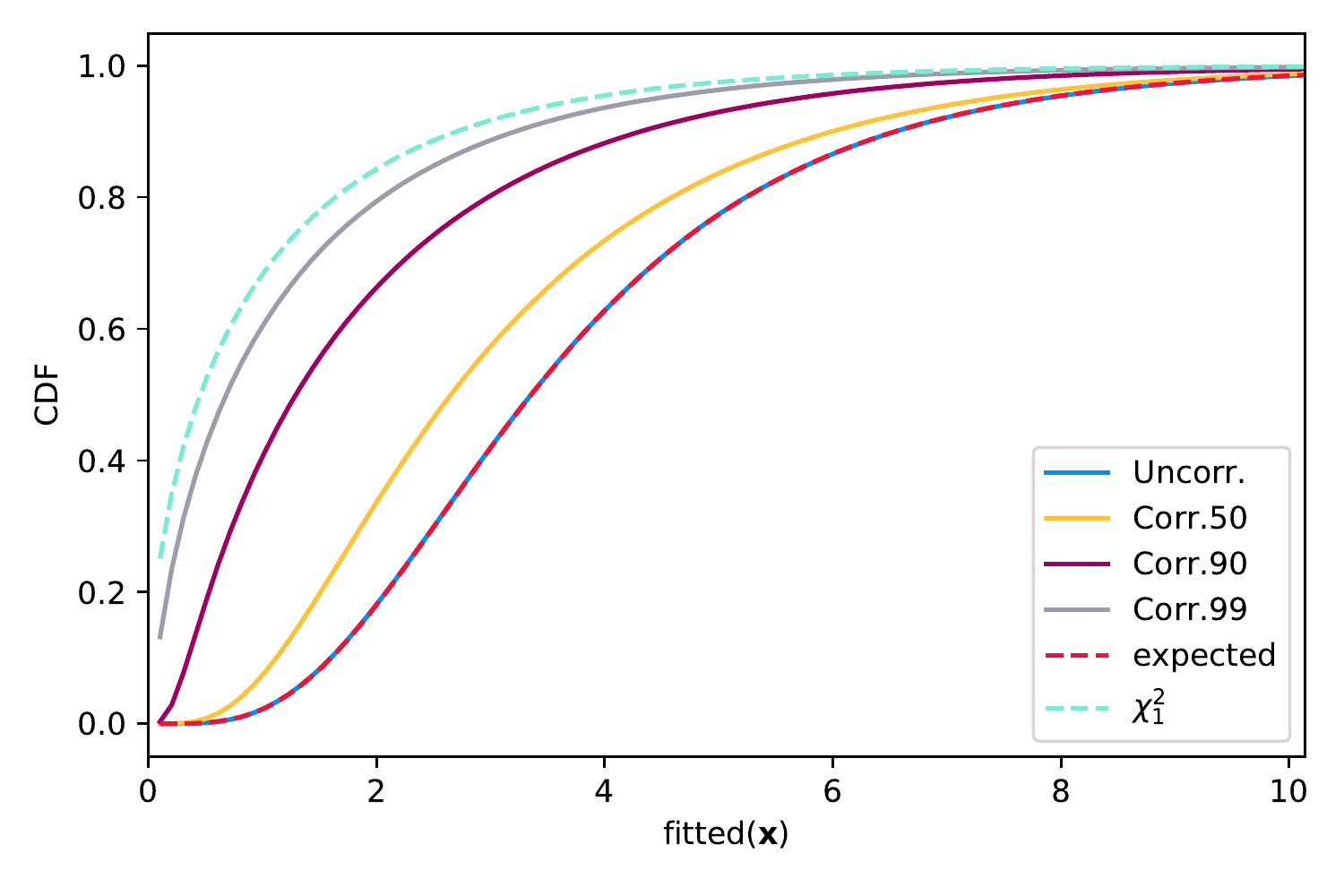}
    \includegraphics[width=0.49\textwidth]{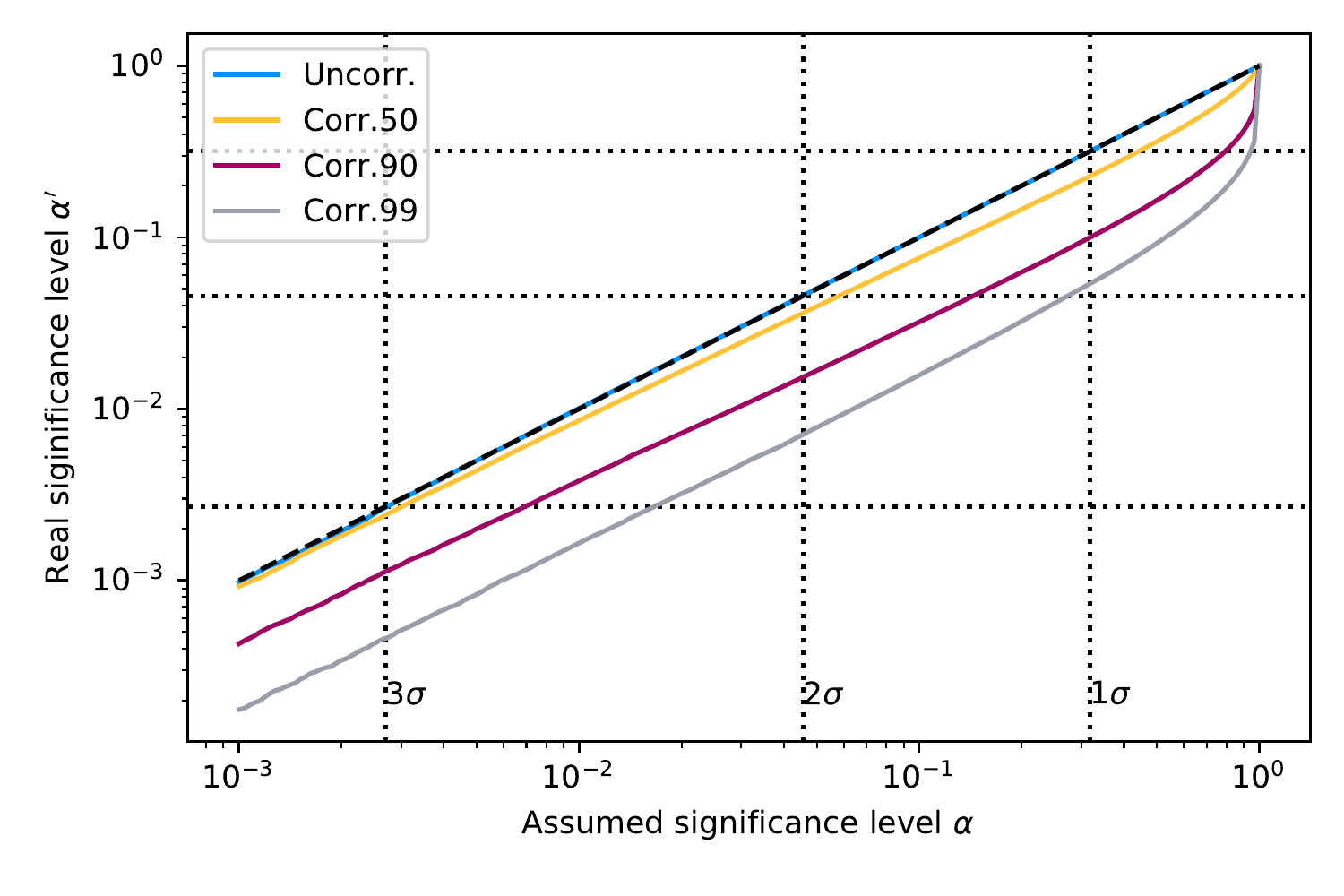}
    \caption{\label{fig:fitted}%
        CDFs (left) for the ``fitted'' test statistic for different levels of correlations in the data.
        When using the uncorrelated CDF to calculate the assumed significance level (or p-value) of a value of the statistic, the actual level will differ from the assumption depending on the correlations (right).
        In the presence of correlations, the real significance is consistently higher (the significance level is lower) than the assumption. This means the uncertainties are overestimated and the statistic behaves conservatively. As the correlations increase, the distribution of the fitted test statistic approaches the $\chi^2_1$ distribution.
    }
\end{figure*}

\section{Asymptotically invariant test statistics}

The fitted test statistic described above is ``safe'' to use in the sense that it is always conservative.
Unfortunately it gets more and more conservative with increasing correlations in the data.
It would be advantageous if the test statistic was exact at all levels of correlations,
or at least at both no correlations, and (in the limit of) 100\% correlations.
To achieve the latter, it is useful to view the problem in the ``CDF space'' of the data points.

Instead of the distribution of $\bm{\Delta}$, let us consider the CDFs of the squares of the \emph{single} variables:
\begin{equation}\label{eq:cdf-space}
    y_i = F_{\chi^2_1}(\Delta_i^2/S_{ii})\text{,}
\end{equation}
where $F_{\chi^2_1}$ is the CDF of $\Delta^2_i/S_{ii}$,
since we assume $\Delta_i$ to be normal distributed with a variance of $S_{ii}$.
Since $y$ is a function of a random variable, it is itself a random variable.
Also, by definition, $y_i$ is uniformly distributed between 0 and 1:
\begin{equation}
    y_i \sim \mathrm{U}(0,1)\text{.}
\end{equation}
This is true irrespective of the possible correlations between the data points,
as long as the marginal distribution of each single data point is known.
In fact, the single data points do not have to be normal distributed.
If they follow a different (but known) distribution, its CDF can be substituted in \autoref{eq:cdf-space}.

If and only if the different variables are independently distributed,
the combined probability density of all $y_i$, $f_y(\bm{y})$ will also be uniform within the N-cube defined by the N unit vectors:
\begin{equation}
    f_y(\bm{y}) =
    \begin{cases}
        1 \qq{if} 0 \le y_i \le 1 \quad \forall i \\
        0 \qq{else.}
    \end{cases}
\end{equation}
This follows from simply multiplying the PDFs of the single variables $y_i$.

If, on the other hand, the variables are perfectly correlated,
the values of all $y_i$ will be identical in each random sampling.
This means the \emph{combined} PDF must be zero wherever the $y_i$ are not identical.
In this case, the combined probability density will be a delta function that concentrates all probability on the main diagonal of the hypercube:
\begin{equation}
    f'_y(\bm{y}) =
    \begin{cases}
        \prod_{i=1}^{N-1} \delta(y_i - y_{i+1}) \qq{if} 0 \le y_1 \le 1 \\
        0 \qq{else}
    \end{cases}
\end{equation}
Note that this does not affect the marginal distributions of the single variables.
Marginalising out all but one variable leads again to a uniform distribution in that variable.
If we can define a function $z(\bm{y})$ that is identically distributed under both assumptions,
we can use it to define a test statistic that is exact in both cases.

\begin{figure*}
    \centering
    \begin{subfigure}{0.49\textwidth}
        \includegraphics[width=\textwidth]{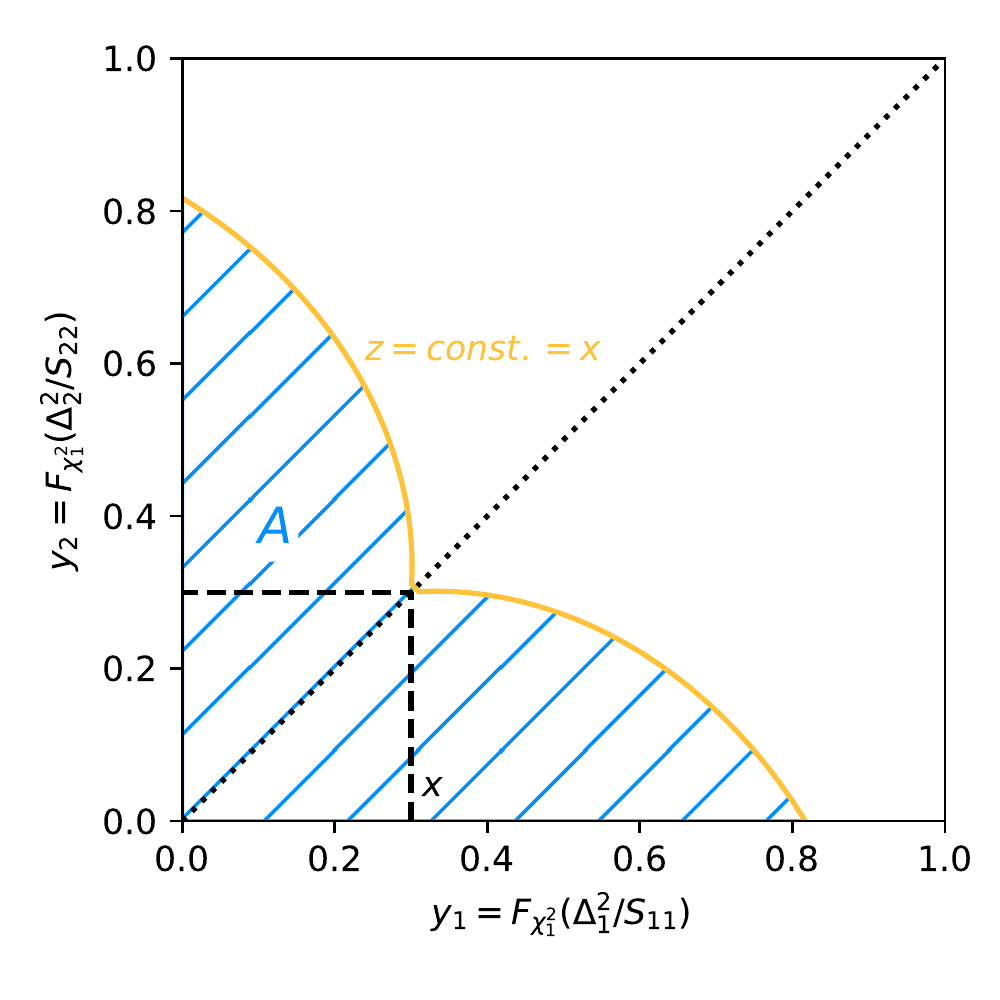}
        \caption{\label{fig:cdf-space}Illustration}
    \end{subfigure}
    \begin{subfigure}{0.49\textwidth}
        \includegraphics[width=\textwidth]{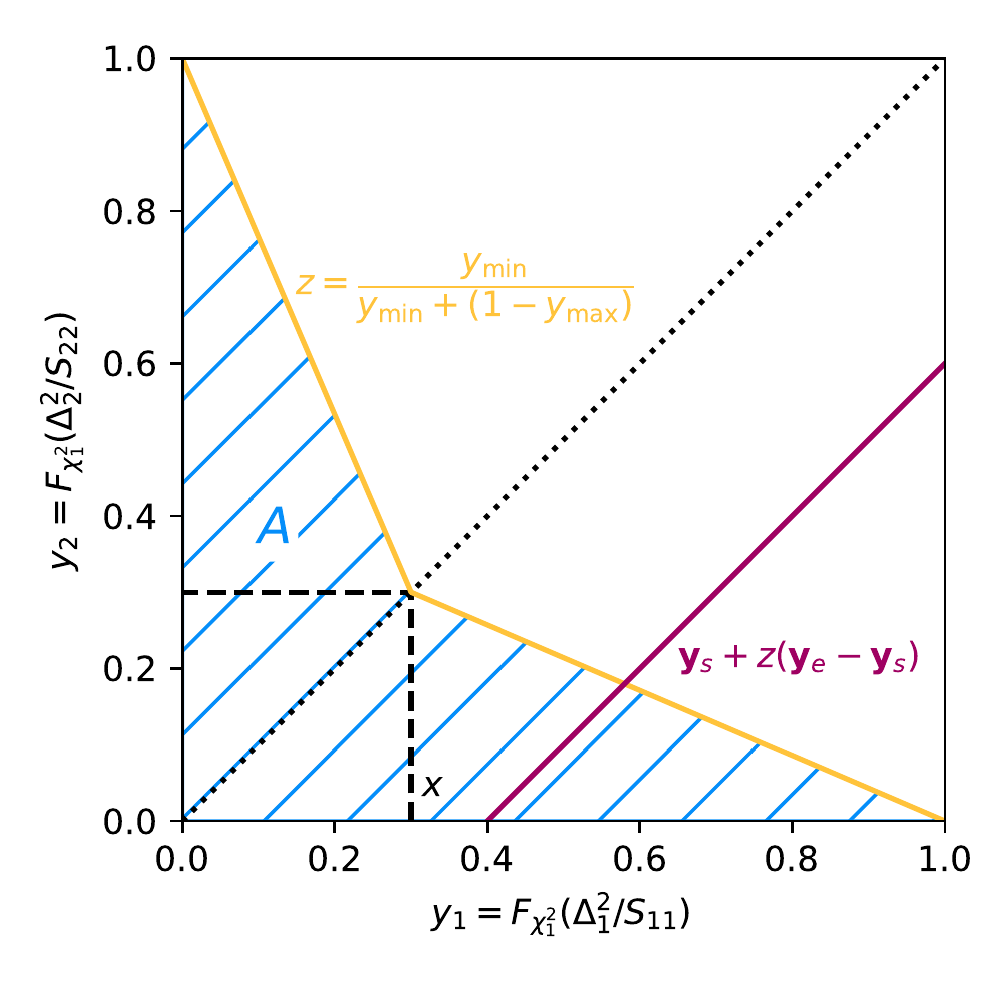}
        \caption{\label{fig:cdf-inv1}Invariant 1}
    \end{subfigure}
    \begin{subfigure}{0.49\textwidth}
        \includegraphics[width=\textwidth]{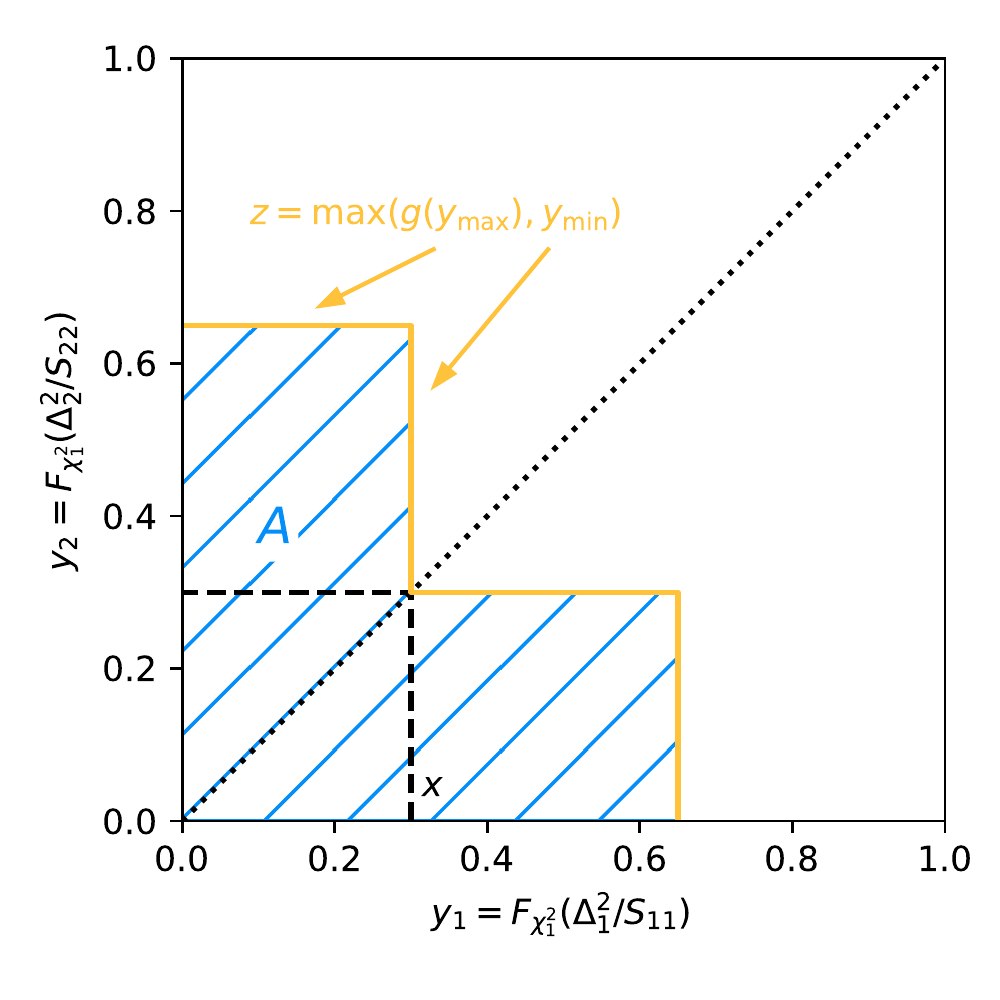}
        \caption{\label{fig:cdf-inv2}Invariant 2}
    \end{subfigure}
    \begin{subfigure}{0.49\textwidth}
        \includegraphics[width=\textwidth]{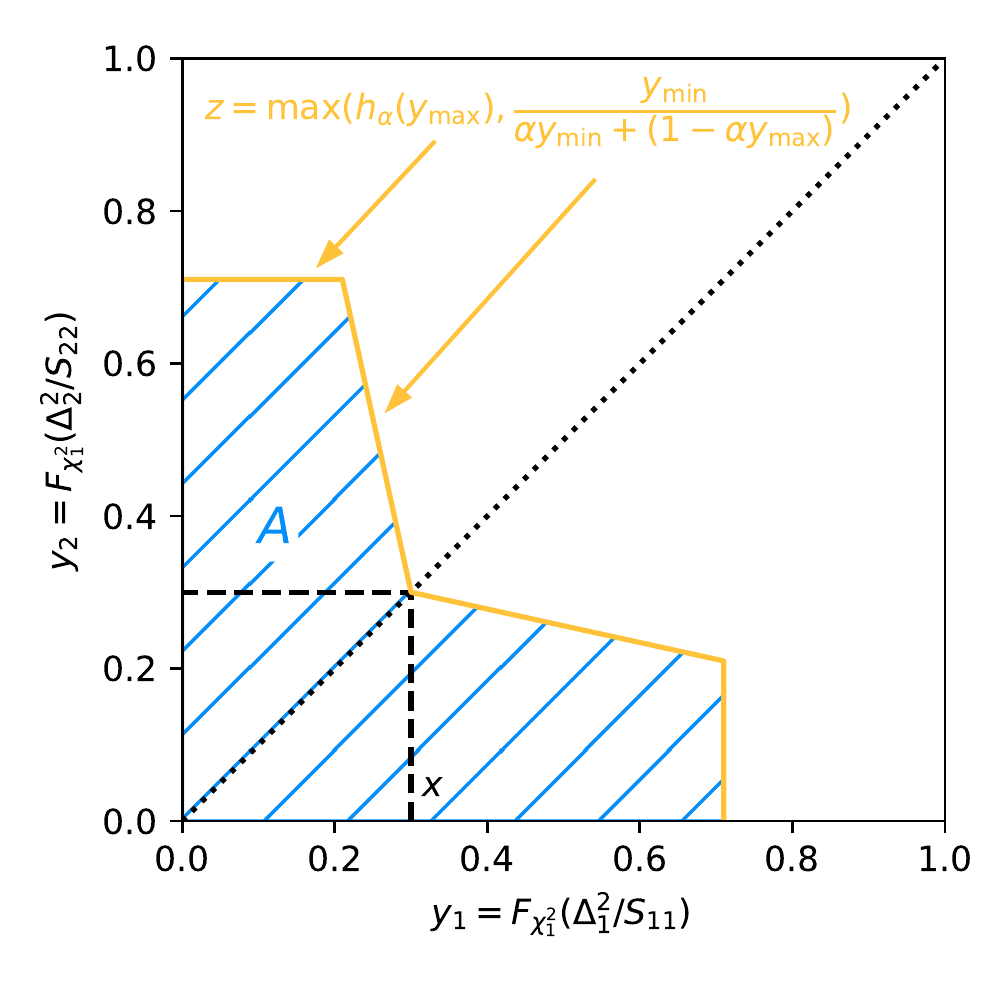}
        \caption{\label{fig:cdf-inv3}Invariant 3}
    \end{subfigure}
    \caption{Illustration of the test statistics in CDF space.
    For $z$ to be identically distributed at both no correlations (PDF of $\bm{y}$ is uniform over square) and at $100\%$ correlations (PDF of $\bm{y}$ is uniform along diagonal),
    the area $A$ enclosed by the implicit function $z = const.$ must be equal to the position $x$ where the function crosses the diagonal.
    }
\end{figure*}

Let us demand that a low value of $z$ indicates a good agreement between data and model, while high values indicate tension between the two.
Within the N-dimensional hypercube, this means that $z$ should be low towards the corner at $\bm{y} = \bm{0}$ and increase towards the corner at $\bm{y} = \bm{1}$.
For $z$ to be identically distributed with no correlations and with $100\%$ correlations,
the surface defined by the implicit function $z(\bm{y}) = z' = \mathrm{const.}$ must enclose the same amount of probability in both cases,
as this defines the CDF of $z$:
\begin{equation}
    F_z(z') = \int_{z(\bm{y}) \le z'} f^{(\prime)}_y(\bm{y}) \dd[N]\bm{y}
\end{equation}
In the case of no correlation, this is the volume $A$ of the part of the N-cube enclosed by the implicit function.
In the case of perfect correlation, it is equal to the single (identical) $y_i$ coordinates where the surface intersects with the diagonal.
\autoref{fig:cdf-space} illustrates this in the case of two dimensions.

Let us call this coordinate $x$ and let us also demand that the function $z(\bm{y}) = x$ at that point.
We then get the following condition:
\begin{equation}\label{eq:condition}
    A = \int_{z(\bm{y}) \le x} \dd[N]\bm{y} \overset{!}{=} x\text{,}
\end{equation}
where the integral is understood to be confined to the inside of the N-cube.
The challenge is now to find functions $z(\bm{y})$ which fulfil this condition for any number of dimensions.

\subsection{Invariant 1}

The simplest way to fulfil \autoref{eq:condition} in two dimensions is to draw straight lines from the points on the diagonal to the "off-diagonal" corners of the square, as shown in \autoref{fig:cdf-inv1}.
We can easily calculate the value of $z(\bm{y})$ for any given point,
as each point can be seen as lying on a straight diagonal line starting at the lower or left edge of the square ($\bm{y}_s$) and ending on the right or top edge ($\bm{y}_e$).
The fractional distance along this line is the desired $z$ and evaluates to
\begin{equation}\label{eq:z1}
    z(\bm{y}) = \frac{y_{\min}}{y_{\min} + (1 - y_{\max})} \text{,}
\end{equation}
where $y_{\min/\max}$ are the minimum and maximum of the elements of $\bm{y}$ respectively.
This also directly applies to the N-dimensional case without change.

Now, it would be possible to use $z$ as the test statistic directly when used on its own.
When the data is intended to be used in conjunction with other data sets though,
e.g. in a global fit, it is useful to use a test statistic that is (approximately) chi-square distributed.
To this end, we can simply apply another function to $z$ which is chosen so that the result is chi-square distributed if $z$ is uniformly distributed.
This function is just the inverse of the CDF of the chi-square distribution $\overset{_{-1}}{F_{\chi^2_1}}$.
Finally we get the first of the ``invariant'' test statistics:
\begin{equation}\label{eq:invariant}
    \invariant_1(\bm\Delta\,|\,\bm{s}) = \overset{_{-1}}{F_{\chi^2_1}}\qty( z\qty( F_{\chi^2_1}\qty(\Delta_i/\sqrt{S_{ii}}),\dots) ) \text{,}
\end{equation}
with $z$ as defined in \autoref{eq:z1}.

Note that we could have chosen a different number of degrees of freedom for the transformation back to a chi-square distribution.
This makes no difference when using the test statistic on a single data set alone, but it changes the relative weight a data set has in a global fit with other data when computing the total chi-square.
If one is confident that the correlations in the data set are weak, it might be better to use the actual number of data points.
In the presence of medium to strong correlations it could be argued though that there is actually less information in the data than the number of data points suggests,
or rather, we are losing ``degrees of freedom'' by having to make up for the missing information of the covariance parameters.

\begin{figure*}
    \centering
    \includegraphics[width=0.49\textwidth]{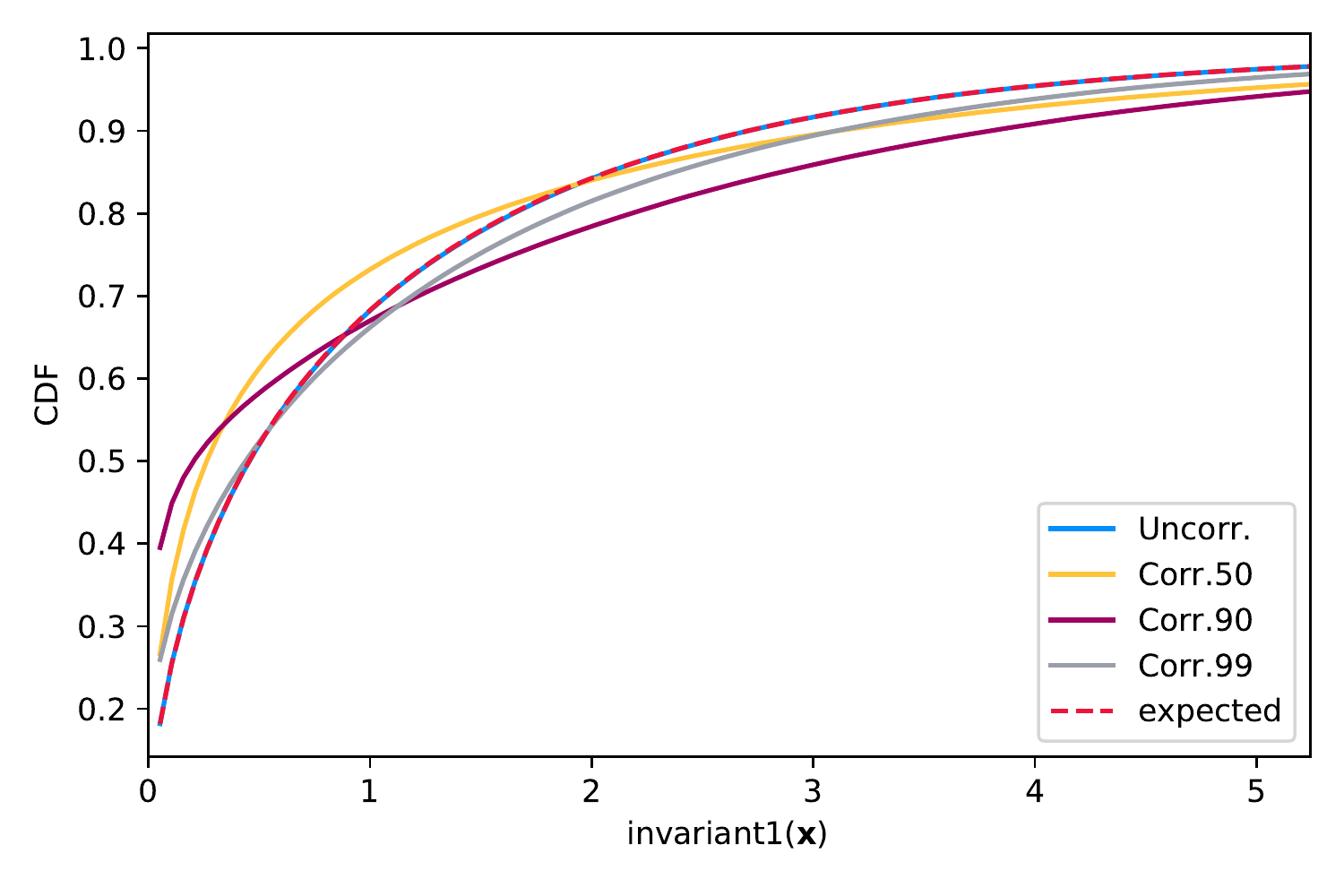}
    \includegraphics[width=0.49\textwidth]{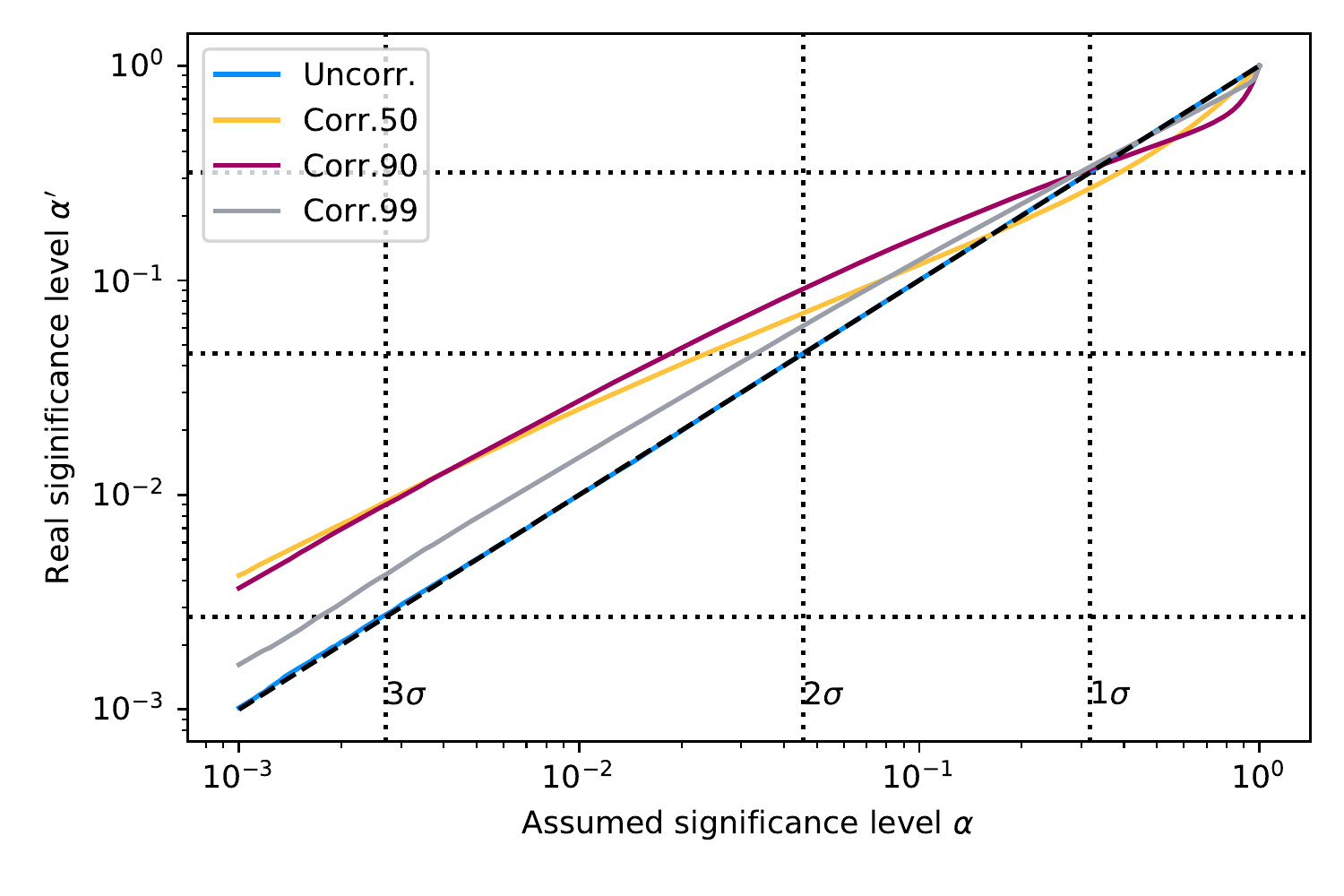}
    \caption{\label{fig:invariant1}%
        CDFs (left) for the ``invariant~1'' test statistic for different levels of correlations in the data.
        When using the uncorrelated CDF to calculate the assumed significance level (or p-value) of a value of the statistic, the actual level will differ from the assumption depending on the correlations (right).
        The effect of the correlations is weaker than for the naive test statistic, but it is not consistently conservative as the fitted one. As the correlations increase, the distribution of the test statistic approaches the uncorrelated expectation again.
    }
\end{figure*}

\autoref{fig:invariant1} shows the performance of the test statistic.
It does deviate from being exact in the presence of correlations,
but the deviation peaks at a certain level and from then on it get more exact again when the correlations are further increased.
Unfortunately it is not conservative for low significance levels.

\subsection{Invariant 2}

Another simple solution in two dimensions is to add identical rectangles to the two inside-facing sides of the $x^2$ square, as shown in \autoref{fig:cdf-inv2}.
Let $l$ be the length of the added rectangles:
\begin{equation}
    l = \frac{1-x}{2}\text{.}
\end{equation}
Extendend into $N$ dimensions, the shape of $A$ becomes a hypercube $H_1$ with an edge length of $x+l$, minus another hypercube $H_2$ at the inside diagonal corner with an edge length of $l$:
\begin{equation}
    A = (x + l)^N - l^N \overset{!}{=} x \text{.}
\end{equation}
Expressed in the total width of the larger cube $d = x+l$ we get:
\begin{gather}\label{eq:g}
    d^N - l^N \overset{!}{=} d - l \text{.}
\end{gather}
This equation will always have a solution of $l \in (0,d)$ for any $d \in (N^{-1/(N-1)},1)$, and can be solved numerically.
Let $l(d)$ be that solution, so we can define the function $g: [0,1] \rightarrow [0,1]$:
\begin{equation}
    g(d) = \begin{cases}
        0 & \qif* d \le N^{-1/(N-1)} \\
        d - l(d) & \qif* N^{-1/(N-1)} < d < 1 \\
        1 & \qif* d = 1 \text{,}
    \end{cases}
\end{equation}
which can calculate $x$ from a given (possible) edge length $d$ of $H_1$.

To calculate the $z$ of any given point, we can use that every point on the surface of $(H_1 \setminus H_2)$ is either on one of the the ``outer'' faces of $H_1$ (where all coordinates are $>0$), or one of the ``inner'' faces of $H_2$ (where at least one coordinate is $x$).
In the former case, the edge length of $H_1$ is given by the maximum of the $y$-coordinates, while in the latter case the position of the ``inner'' diagonal corner of $H_2$ is given by the minimum of the coordinates.
Since that inner corner is at $y_i = x\ \forall\ i$ by construction,
we can write $z$ as
\begin{equation}
    z(\bm{y}) = \max(g(y_{\max}), y_{\min})\text{.}
\end{equation}
With this $z$ we can then define the $\invariant_2(\bm{\Delta}\,|\,\bm{s})$ test statistic just like in \autoref{eq:invariant}.

\begin{figure*}
    \centering
    \includegraphics[width=0.49\textwidth]{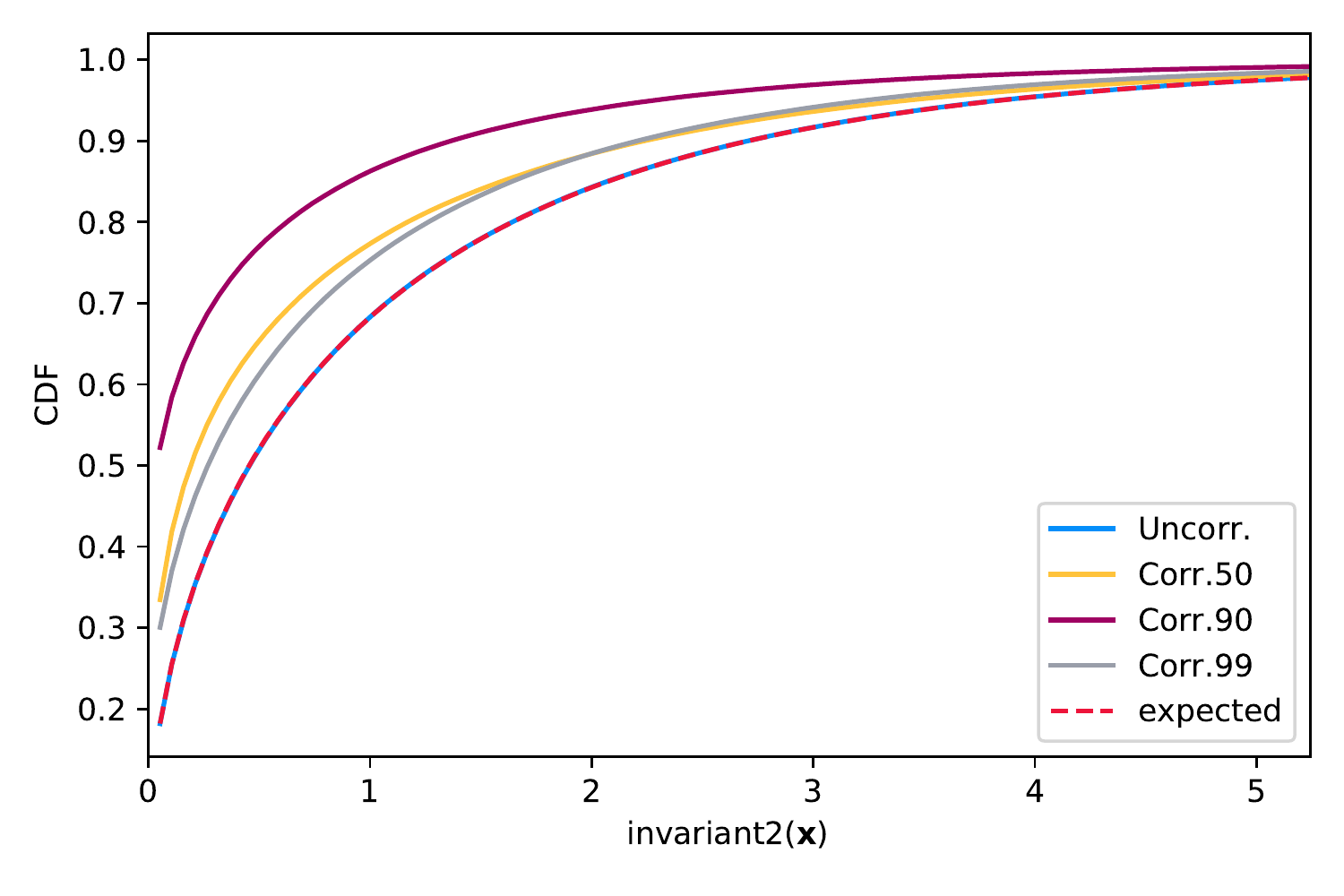}
    \includegraphics[width=0.49\textwidth]{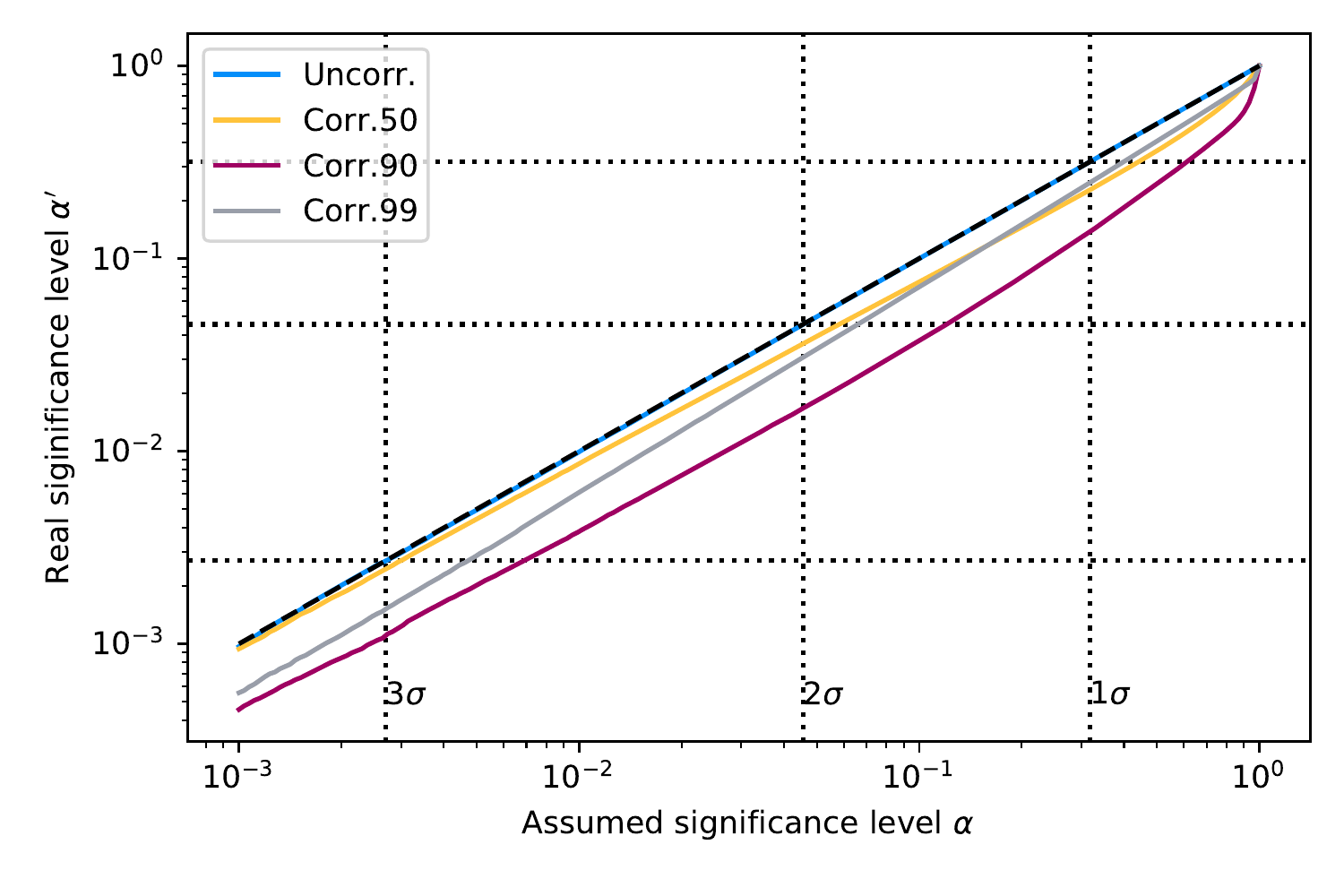}
    \caption{\label{fig:invariant2}%
        CDFs (left) for the ``invariant~2'' test statistic for different levels of correlations in the data.
        When using the uncorrelated CDF to calculate the assumed significance level (or p-value) of a value of the statistic, the actual level will differ from the assumption depending on the correlations (right).
        Like the fitted test statistic, this one is consistently conservative. As the correlations increase, the distribution of the test statistic approaches the uncorrelated expectation again.
    }
\end{figure*}

Its performance is shown in \autoref{fig:invariant2}.
It is conservative for all strengths of correlation and significance levels,
and shows the expected limit of exactness at no and very strong correlations.

\subsection{Invariant 3}

Finally, \autoref{fig:cdf-inv3} shows an intermediate shape between ``invariant~1'' and ``invariant~2''.
It can be interpreted as the shape of ``invariant~1'', but instead of connecting the diagonal to the off-diagonal corners of the square, it is connected to the respective corners of a larger square with edge length $1/\alpha$,
with the shape parameter $\alpha \in (0,1)$.
To ensure that the resulting area $A$ is equal to $x$,
it is cut off at the edges of a square with edge length $d$.

In N dimensions, the volume of such a body is:
\begin{equation}
    A = d^N - \frac{(d-x)^N}{(1-\alpha x)^{N-1}} \overset{!}{=} x\text{.}
\end{equation}
This equation has one solution of $x \in (0, d)$ if $(N - \alpha d N + \alpha d) > d^{1-N}$.
Let $x(d)$ be that solution and, like before, we can define a function $h_\alpha : (0,1) \rightarrow (0,1)$:
\begin{equation}
    h_\alpha(d) = \begin{cases}
        0 & \qif*  (N - \alpha d N + \alpha d) \le d^{1-N} \\
        x(d) & \qif* (N - \alpha d N + \alpha d) > d^{1-N} > 1 \\
        1 & \qif* d = 1 \text{.}
    \end{cases}
\end{equation}
This determines the value of $z$ for points on the surface of the hypercube.

The value of $z$ for points on the ``cut-off'' corner can be calculated just like in the case of ``invariant~1''.
Only the scaling of the containing cube needs to be taken into account.
The total $z$ function is then again the maximum of the two values:
\begin{equation}
    z(\bm{y}) = \max\qty(h_\alpha(y_{\max}), \frac{y_{\min}}{\alpha y_{\min} + (1 - \alpha y_{\max})})\text{.}
\end{equation}
The final test statistic ``invariant~3'' is then built according to \autoref{eq:invariant}.
A python implementation of this test statistic is provided in Listing~ \ref{lst:invariant} in the appendix.

This test statistic has one free shape parameter $\alpha$.
It determines the opening angle of the iso-$z$ surface where it meets the main diagonal of the hypercube.
In the 2D~case, this angle is constant at $90^\circ$ for $\alpha \rightarrow 0$,
making it identical to the ``invariant~2'' case.
For $\alpha > 0$, the angle starts at $90^\circ$ at $x = 0$ and then opens up with increasing $x$.
The value of $\alpha$ determines where the angle reaches $180^\circ$:
\begin{equation}
    x_{180^\circ} = \frac{1}{2\alpha}\text{.}
\end{equation}
E.g. for $\alpha = 1$, corresponding to the ``invariant~1'' case, the opening angle is $180^\circ$ at $x = 0.5$.

In the presence of ``medium'' correlations,
the opening angle can give an indication of whether the test statistic is conservative for the corresponding significance level.
An opening angle $< 180^\circ$ means that the surface ``protrudes'' into parts of the CDF space that should ``belong'' to a higher value of $x$,
suggesting a conservative statistic.
Conversely, an opening angle $>180^\circ$ means that the CDF space perpendicular to the diagonal is only covered with higher $x$, suggesting a coverage that is actually lower than the expectation.
At $\alpha = 0.5$, the opening angle is $<180^\circ$ for all $x$ (as $x$ is always $<1$).
This makes that value a conservative choice.

\begin{figure*}
    \centering
    \includegraphics[width=0.49\textwidth]{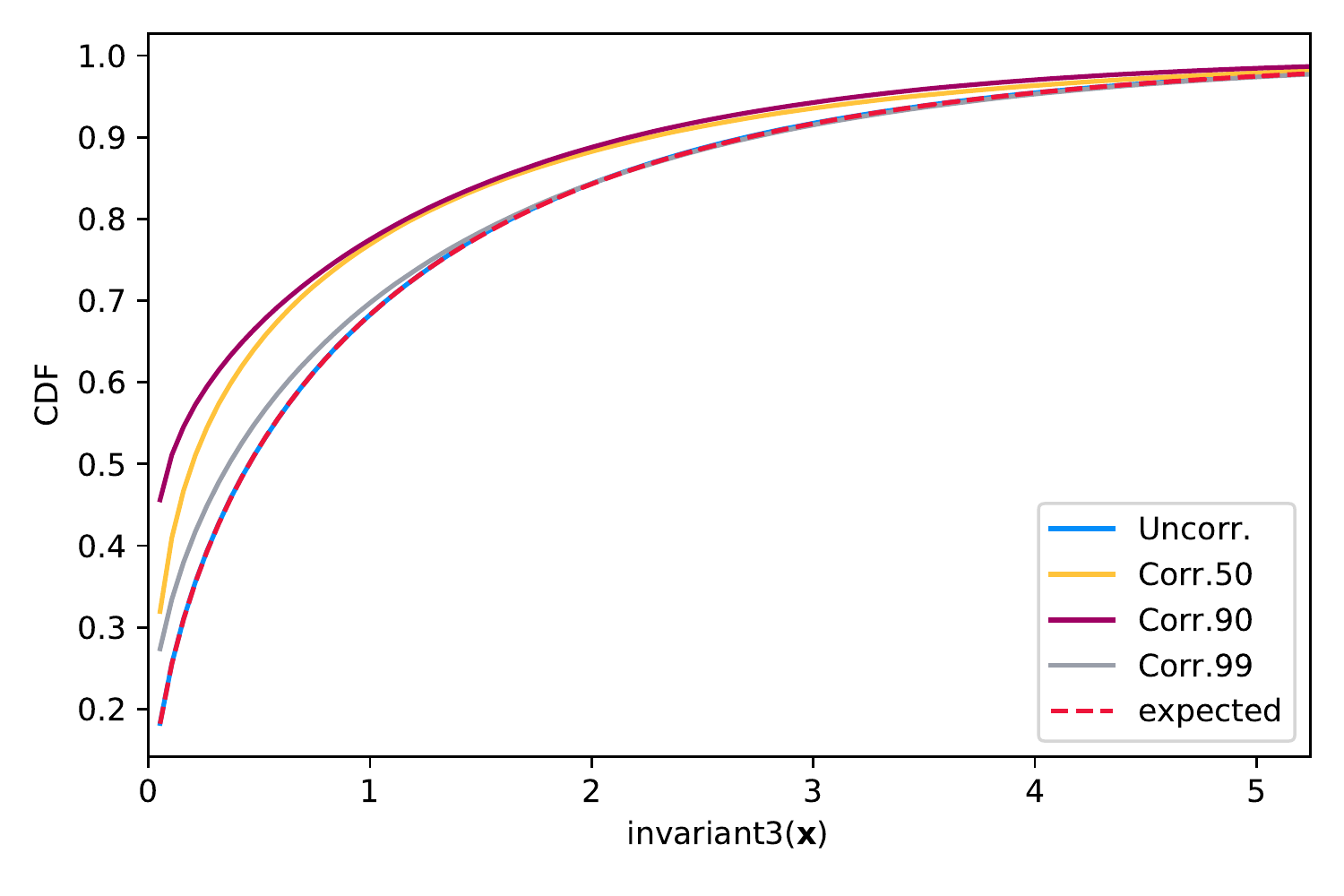}
    \includegraphics[width=0.49\textwidth]{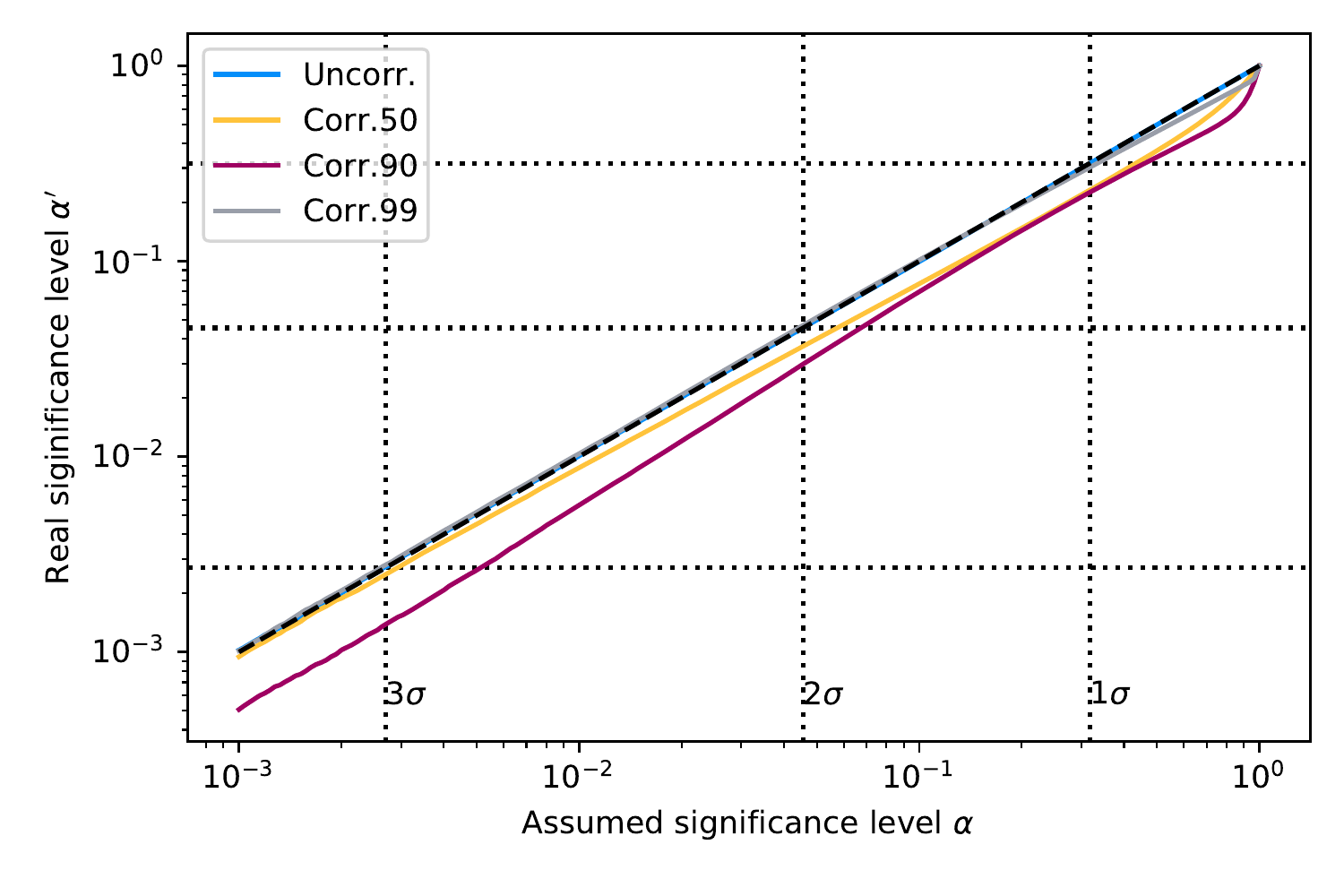}
    \caption{\label{fig:invariant3}%
        CDFs (left) for the ``invariant~3'' test statistic for different levels of correlations in the data.
        When using the uncorrelated CDF to calculate the assumed significance level (or p-value) of a value of the statistic, the actual level will differ from the assumption depending on the correlations (right).
        The behaviour of this statistic depends on the parameter $\alpha$. For $\alpha = 1$ it is identical to the ``invariant~1'' statistic, while for $\alpha \rightarrow 0$ it approaches ``invariant ~''. Shown here is $\alpha = 2/3$.
    }
\end{figure*}

There is room for an more aggressive choice of $\alpha$ though.
\autoref{fig:invariant3} shows the performance of the ``invariant~3'' test statistic with $\alpha = 2/3$.
Out of all considered test statistics, it shows the smallest deviations from exactness,
and it is conservative for the considered significance levels and correlations.

\section{Comparison}
\label{sec:comparison}

It is quite clear that among the considered test statistics,
``invariant~3'' performs the most consistent under many different levels of correlations.
It suffers from a kind of arbitrariness though when trying to combine it in larger fits with other data sets.
The transformation to a chi-square distribution with one degree of freedom is a conservative choice that allows it to be combined in least-squares or likelihood fits.
Aside from the case of perfect correlations, it will under-estimate the amount of information compared to the other data-sets though.

The ``fitted'' statistic is quite a bit more conservative and it gets more conservative the stronger the correlations are.
It has the advantage though that it corresponds to an ``actual Mahalanobis distance'' when considering the correlations in the data as nuisance parameters.
It should thus easily be included in least-square fits in combination with other data sets.
Wilks' theorem does \emph{not} hold for it though, except for very strong correlations.
Only then is it distributed like a chi-square with one degree of freedom.
With no correlations present, it follows the ``Bee-square'' distribution.

\begin{figure*}
    \centering
    \includegraphics[width=0.49\textwidth]{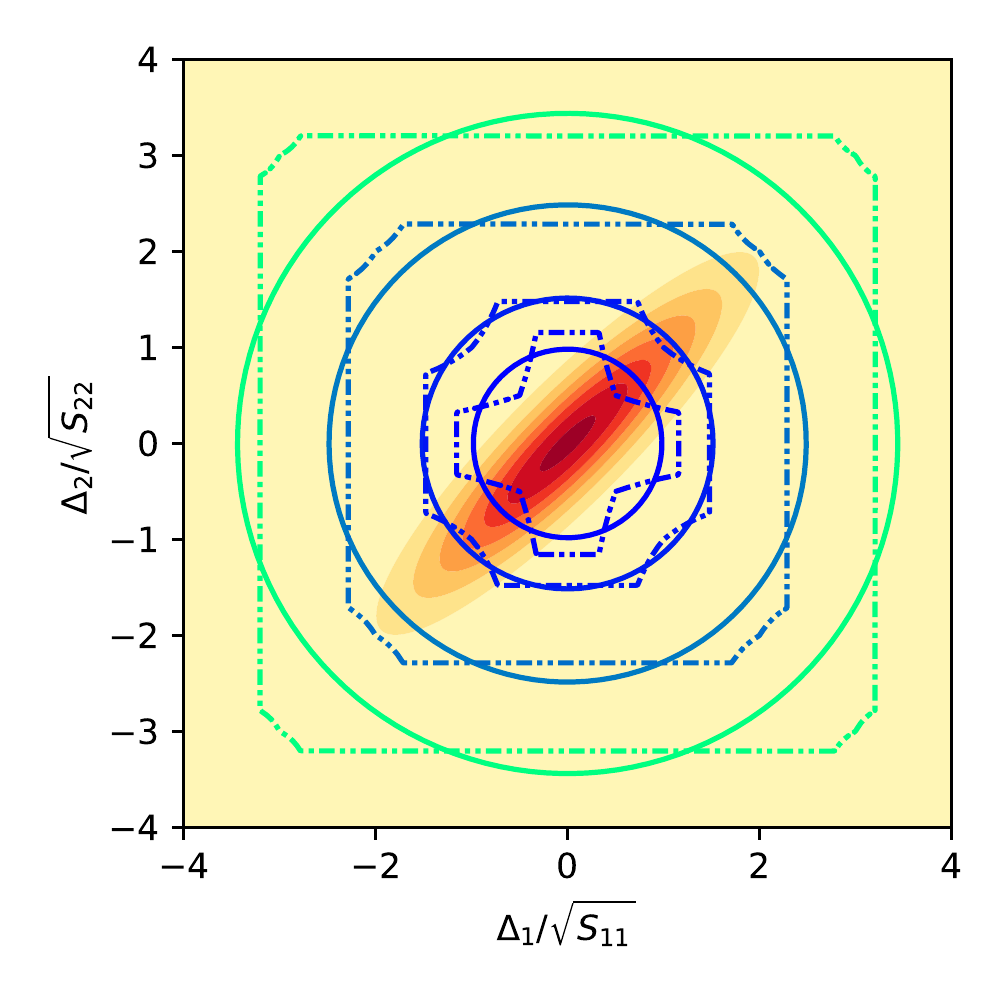}
    \includegraphics[width=0.49\textwidth]{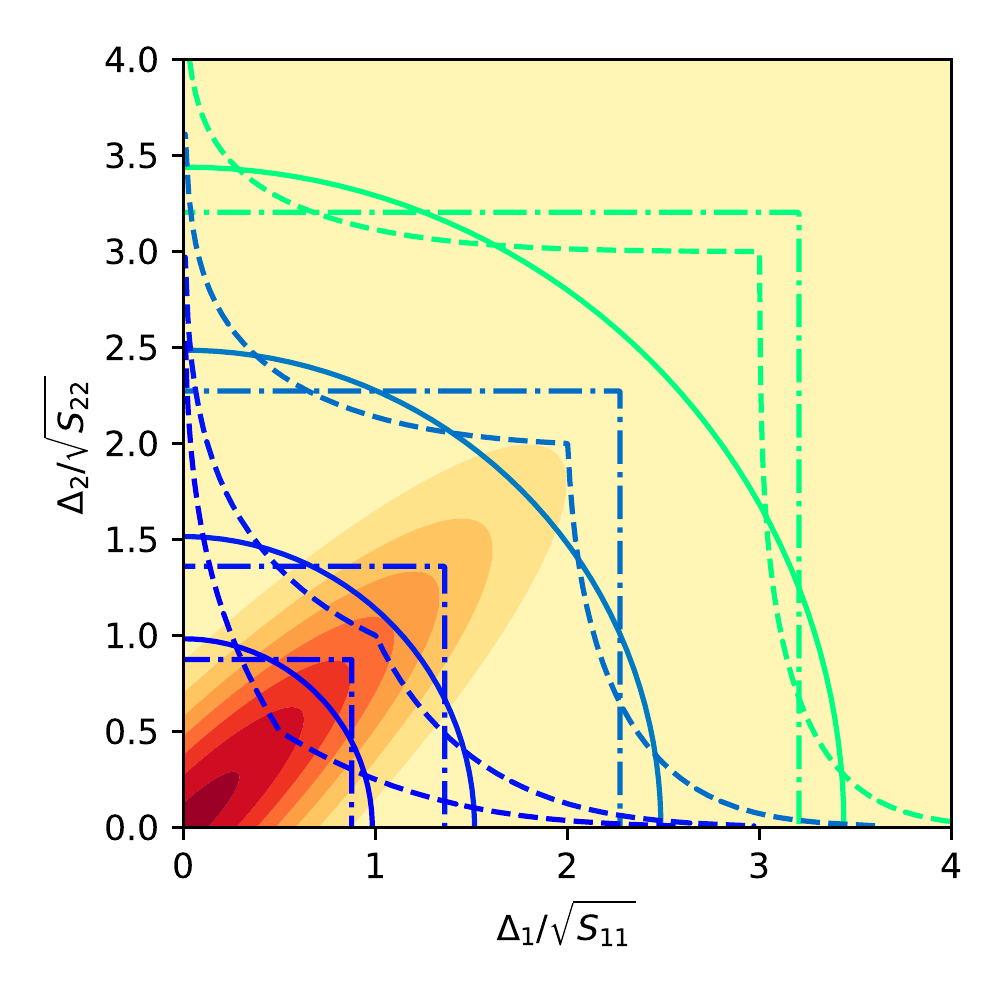}
    \includegraphics[width=0.49\textwidth]{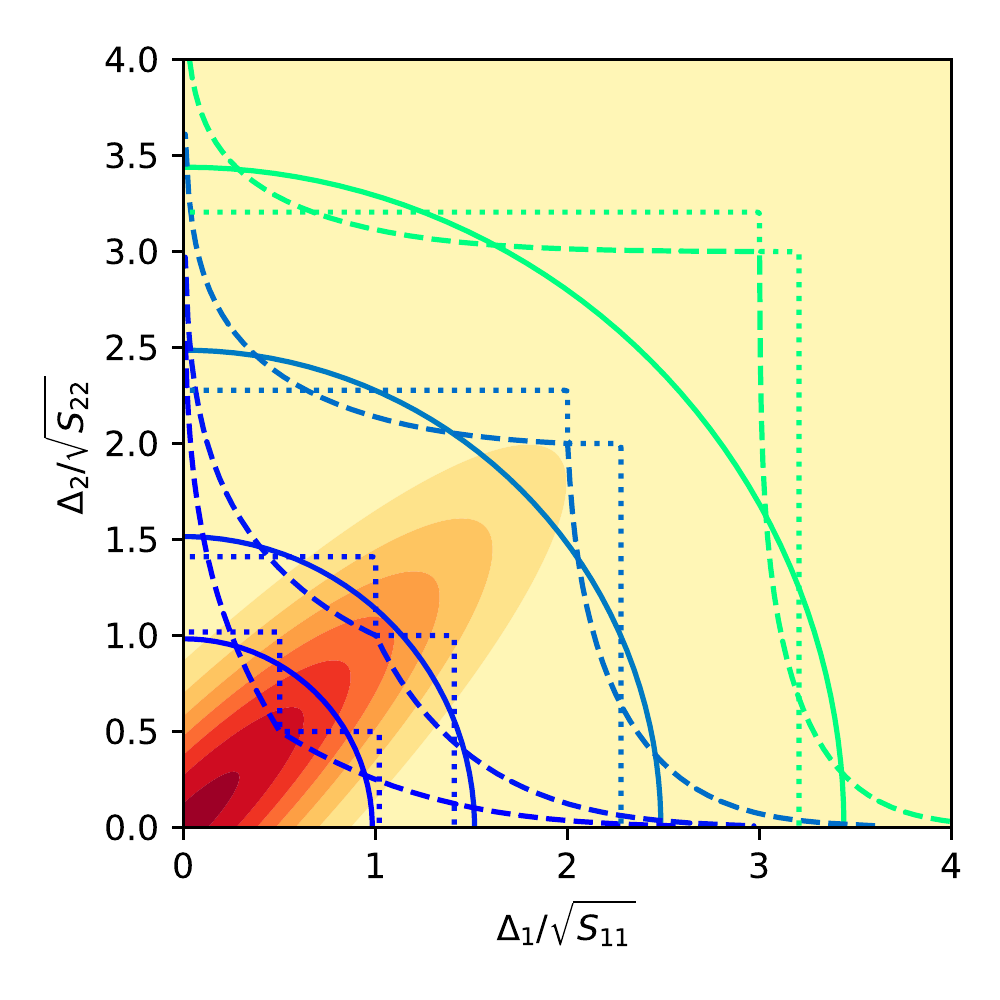}
    \includegraphics[width=0.49\textwidth]{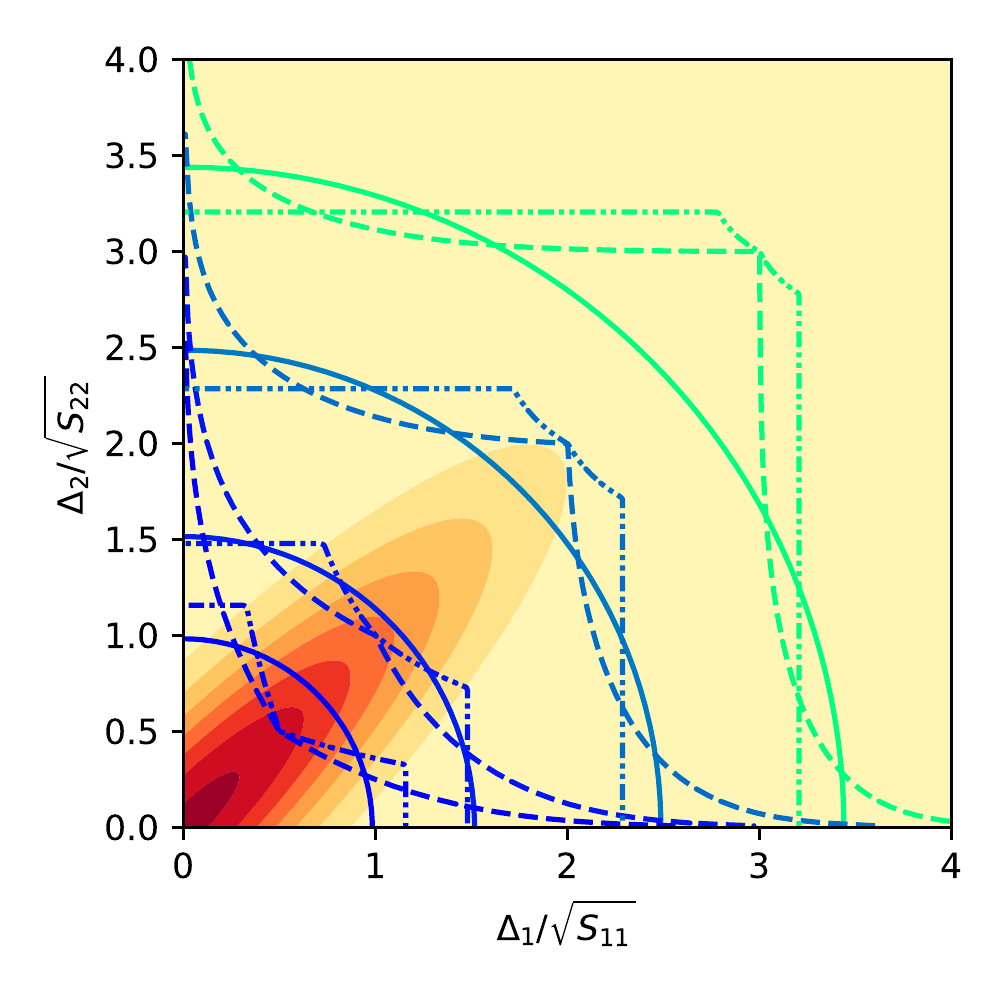}
    \caption{\label{fig:CR}%
    Shape of confidence regions of the different statistis in two dimensions.
    The confidence levels correspond to a $0.5\sigma$, $1\sigma$, $2\sigma$, and $3\sigma$ deviation of a one-dimensional normal distributed variable.
    The shown test statistics are: top left, "naive" (solid) and "invariant~3" (dashdotdot); top right, "naive" (solid), "fitted" (dashdot), "invariant~1" (dashed); bottom left, "naive" (solid), "invariant~1" (dashed), "invariant~2" (dotted); bottom right, "naive" (solid), "invariant~1" (dashed), "invariant~3 ($\alpha=2/3$)" (dashdotdot).
    Also shown is the distribution of some multivariate normal data with a variance of 1.0 and a correlation of 0.9.
    }
\end{figure*}

\autoref{fig:CR} shows the shape of the resulting confidence regions for the different test statistics in two dimensions.
The ``invariant'' statistics all have a cross shape at low confidence levels and then get progressively more square.
The confidence regions of the ``invariant~1'' statistic extend all the way to $\pm\infty$ along the variable axes.
This is due to the fact that all lines of the construction in the CDF space go to the edges of the hypercube.
For a normal distributed variable a CDF of 1 means a variable value of $\pm\infty$.
The coverage is still correct in the case of no correlations,
but in principle this means that one would have to include a point in the $1\sigma$ region that is arbitrarily far away on one axis, as long as it is close enough to 0 in the other axis.
This is clearly counter-intuitive and could very well lead to strange behaviour in fits.
Combined with the fact that it is not conservative for all confidence levels,
the ``invariant~1'' statistic should probably not be used.

\begin{figure*}
    \centering
    \includegraphics[width=0.49\textwidth]{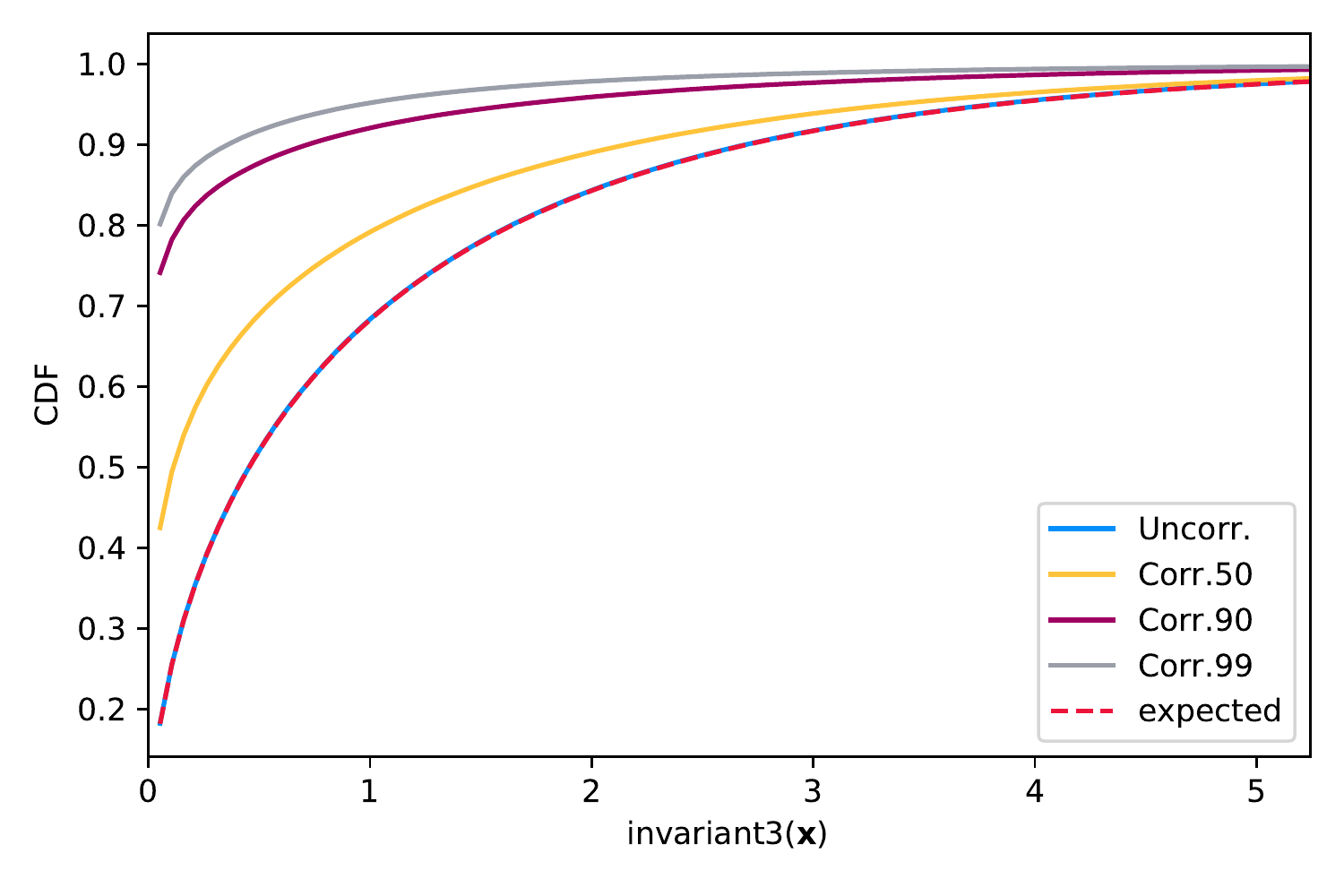}
    \includegraphics[width=0.49\textwidth]{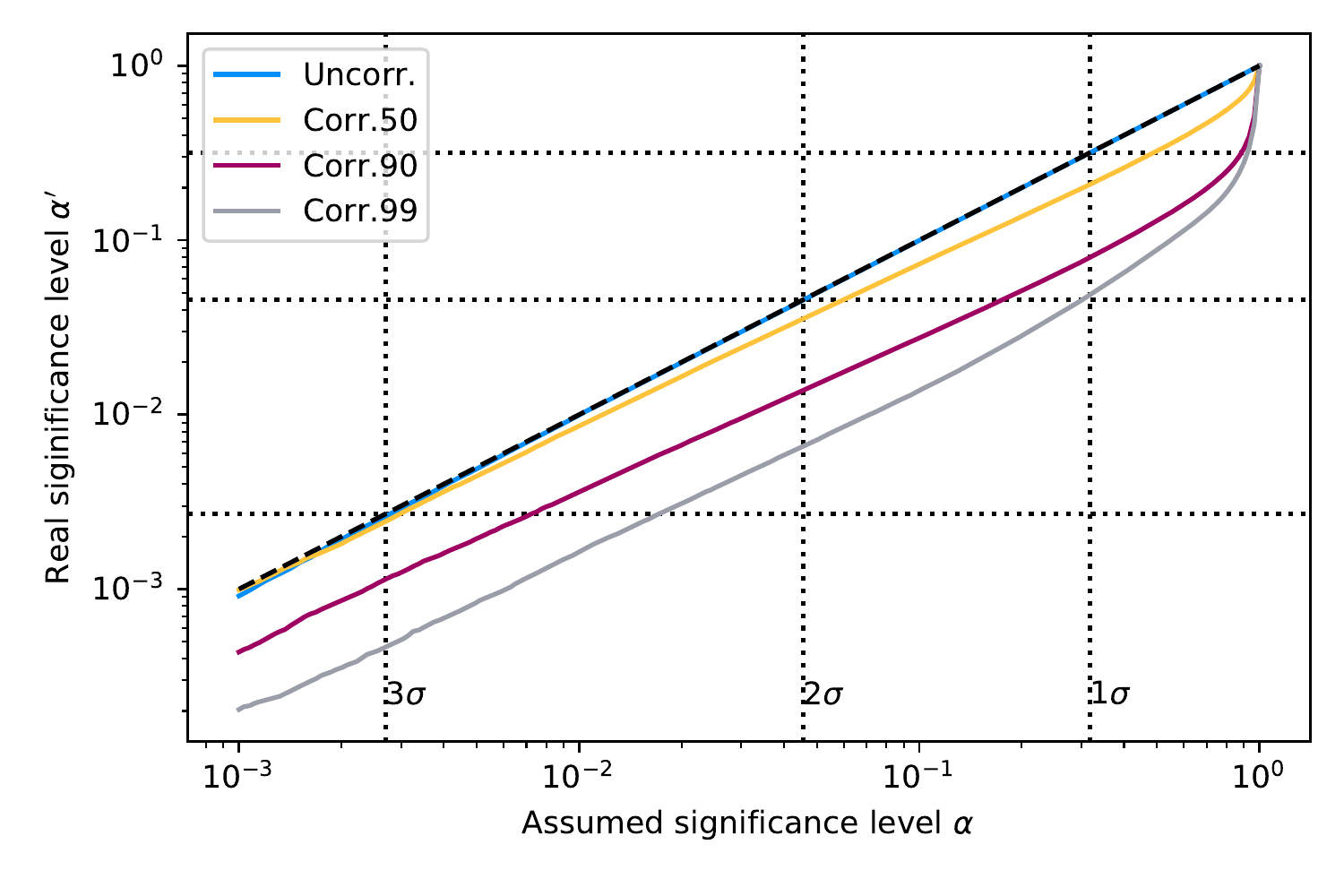}
    \caption{\label{fig:realistic}%
        CDFs (left) for the ``invariant~3'' test statistic for different levels of correlations in data with a more complicated correlation structure (see text). When using the uncorrelated CDF to calculate the significance level (or p-value) of a value of the statistic, the actual level depends on the correlations (right).
    }
\end{figure*}

\begin{figure*}
    \centering
    \includegraphics[width=0.49\textwidth]{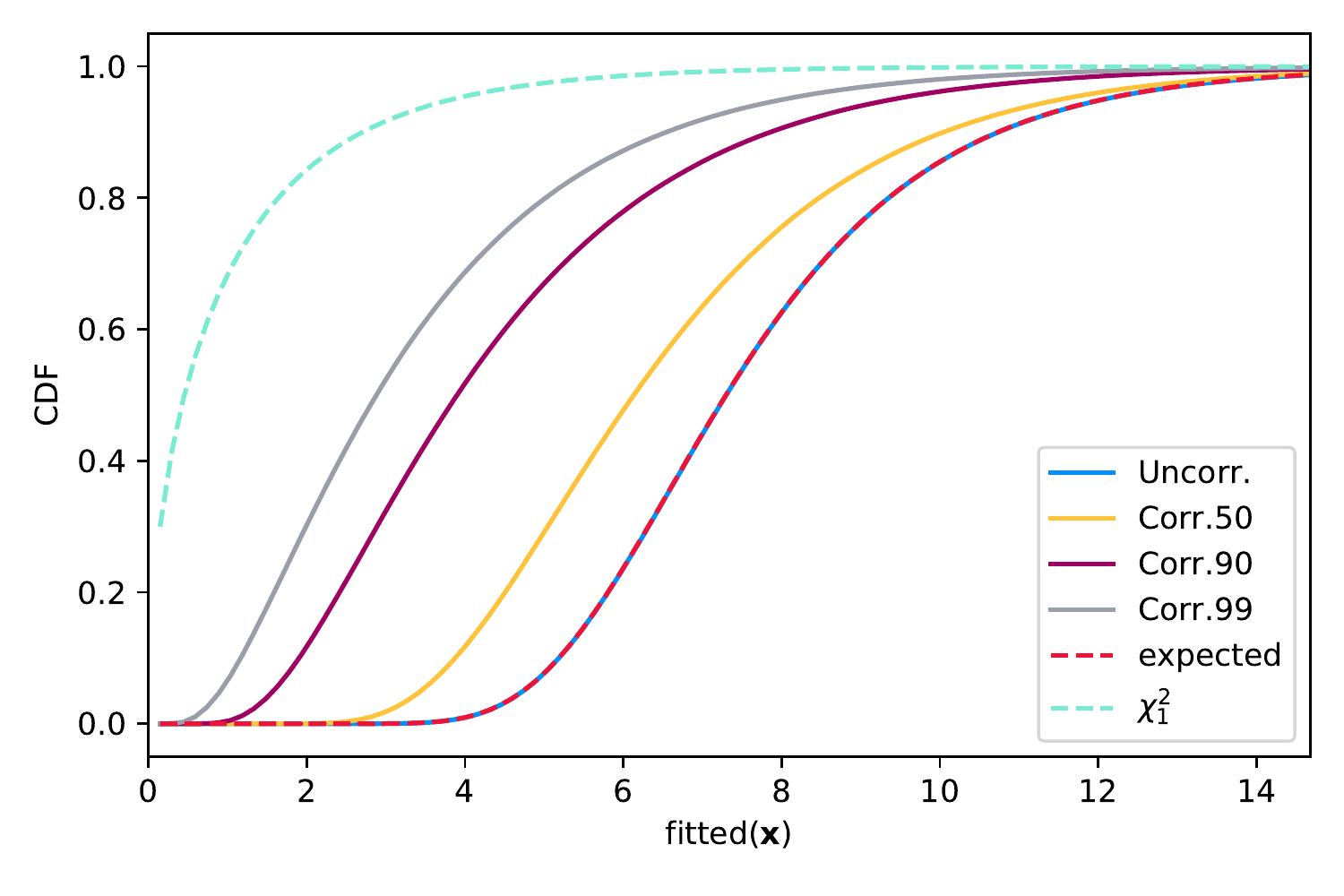}
    \includegraphics[width=0.49\textwidth]{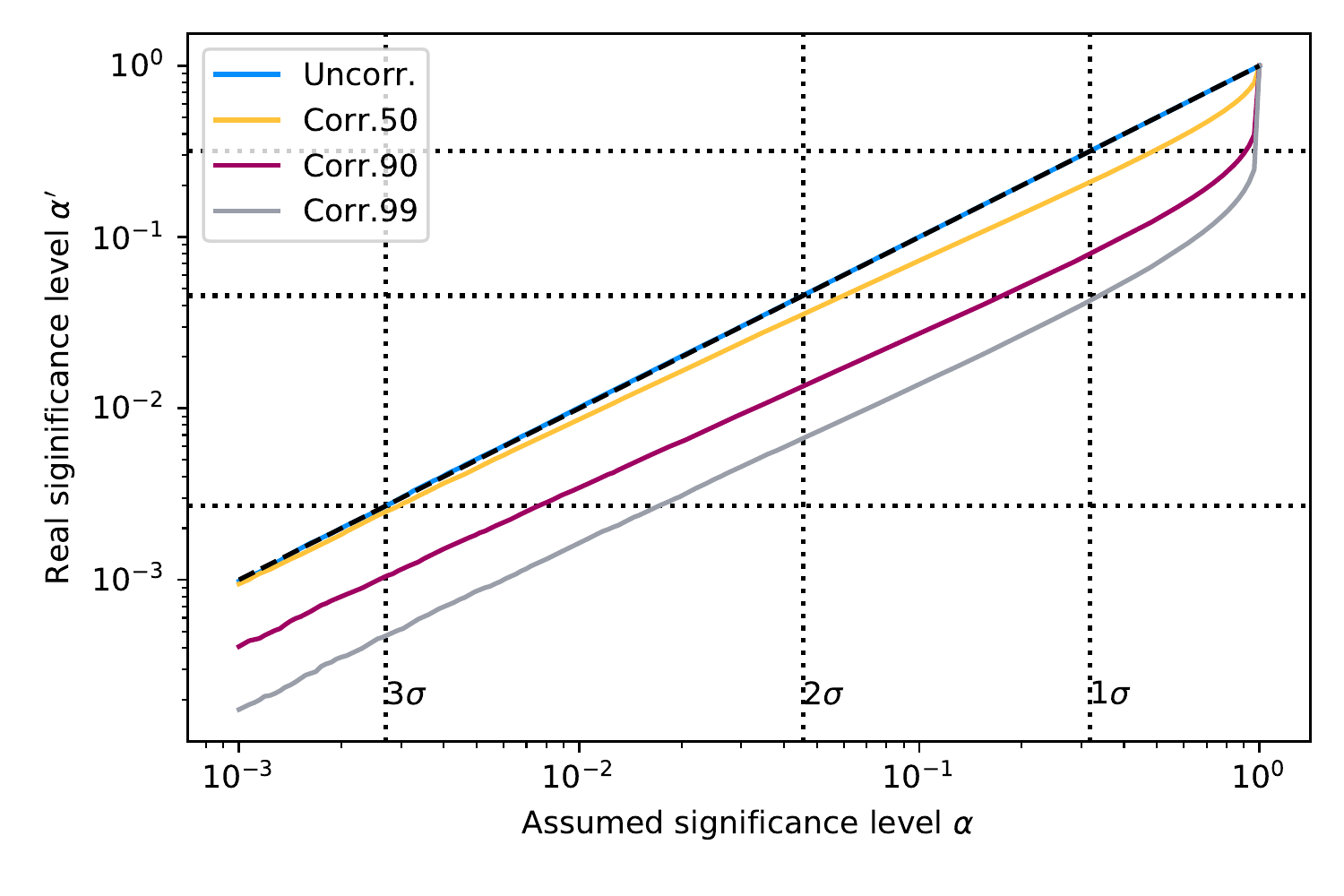}
    \caption{\label{fig:realistic-fitted}%
        CDFs (left) for the ``fitted'' test statistic for different levels of correlations in data with a more complicated correlation structure (see text). When using the uncorrelated CDF to calculate the significance level (or p-value) of a value of the statistic, the actual level depends on the correlations (right).
    }
\end{figure*}

It is worth stressing that the actual coverage behaviour of the test statistics will depend on the actual correlations present in the data sets.
The examples shown here were very simple, with constant correlation coefficients between all data points.
\autoref{fig:realistic} shows the performance of the ``invariant~3'' test statistic for a data set with a more complicated correlation structure.
The data consists of $N = 100$ data points with a covariance matrix of the form
\begin{equation}
    S = \mqty*(
        1       & c_0     & c_1     & \dots  & c_{N-2} \\
        c_0     & 1       & c_0     & \dots  & c_{N-3} \\
        c_1     & c_0     & 1       & \dots  & c_{N-4} \\
        \vdots  & \vdots  & \vdots  & \ddots & \vdots  \\
        c_{N-2} & c_{N-3} & c_{N-4} & \cdots & 1
    )\text{.}
\end{equation}
Here $c_0$ is the maximum correlation as specified in the plot labels,
and $c_{N-2} = -c_0 / 2$.
All intermediate $c_i$ are linearly equidistant, so $c_{i} - c_{i+1} = const$.
The correlations are thus stronger for ``neighbouring'' data points and there is some negative correlation as well.
Since the total covariance never reaches the limit of ``perfect correlation'',
the limiting exactness is not as efficient in this data set and the performance is closer to that of the ``fitted'' test statistic as shown in \autoref{fig:realistic-fitted}.
Since the ``fitted'' test statistic is considerably easier to calculate,
it might be worth choosing it over an ``invariant'' one,
depending on the reasonably expected correlations and computation requirements.

\section{Application to real neutrino cross-section data}
\label{sec:application}

As a test, let us apply the ``invariant~3'' test statistic to some real data and compare its results with the ``naive'' approach.
We will use the double-differential, charged-current quasi-elastic cross section measurement of (anti-)muon neutrinos by MiniBooNE\cite{AguilarArevalo2010, AguilarArevalo2013}.
This data is presented in the form of a set of differential cross sections with a ``shape error'' for each bin, plus a relative ``normalisation error'' common to all bins.
The publications provide no information about the correlations between the bins, or between the shape and the normalisation error.

Model predictions for the measurements were generated with NUISANCE\cite{Stowell2017}.
The generators and models considered in these studies are:
\begin{enumerate}
    \item GENIE\cite{Andreopoulos2010} v3.00.06 tune G18\_10a\_02\_11a
    \item GENIE v3.00.06 tune G18\_10b\_00\_000 
    \item NEUT\cite{Hayato2009} v5.4.1
    \item NuWro\cite{Golan2012} v19.02.2
    \item GENIE v3.00.06 with SuSAv2\cite{GonzalezJimenez2014}
\end{enumerate}

\begin{figure*}
    \centering
    \includegraphics[width=\textwidth]{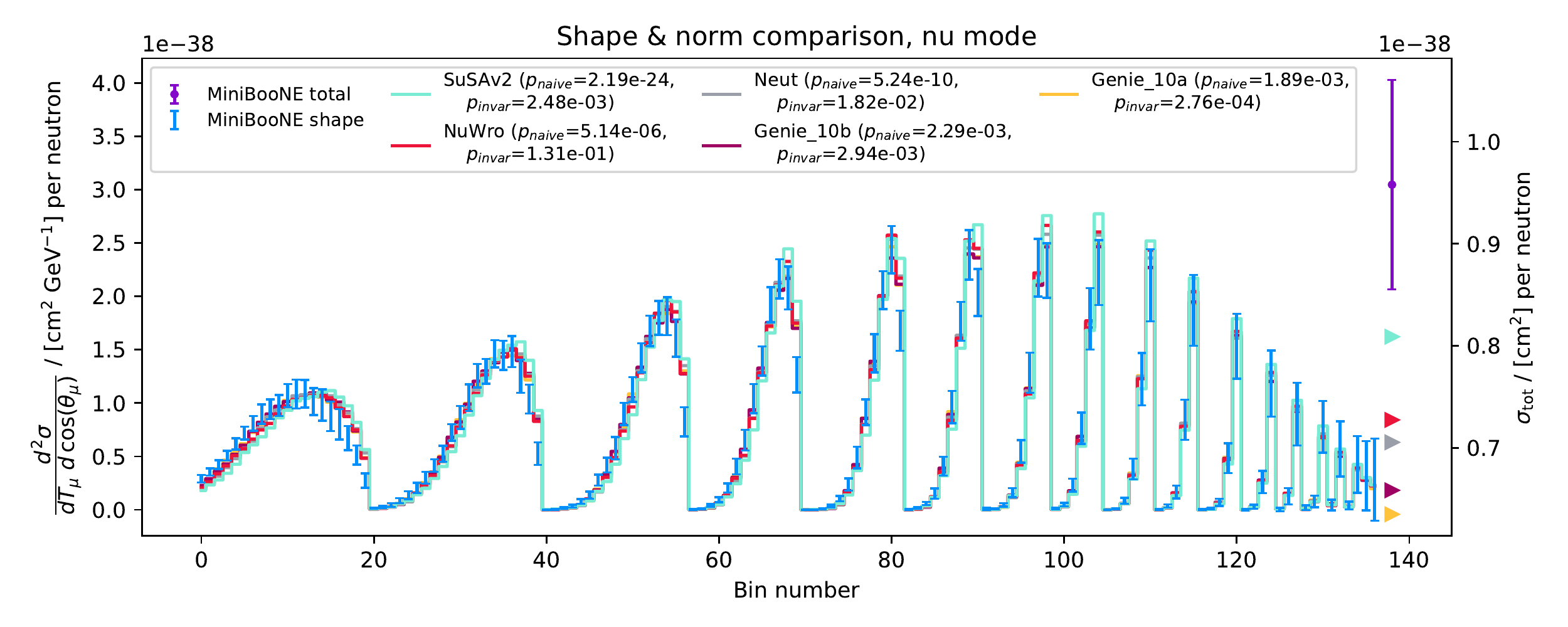}
    \includegraphics[width=\textwidth]{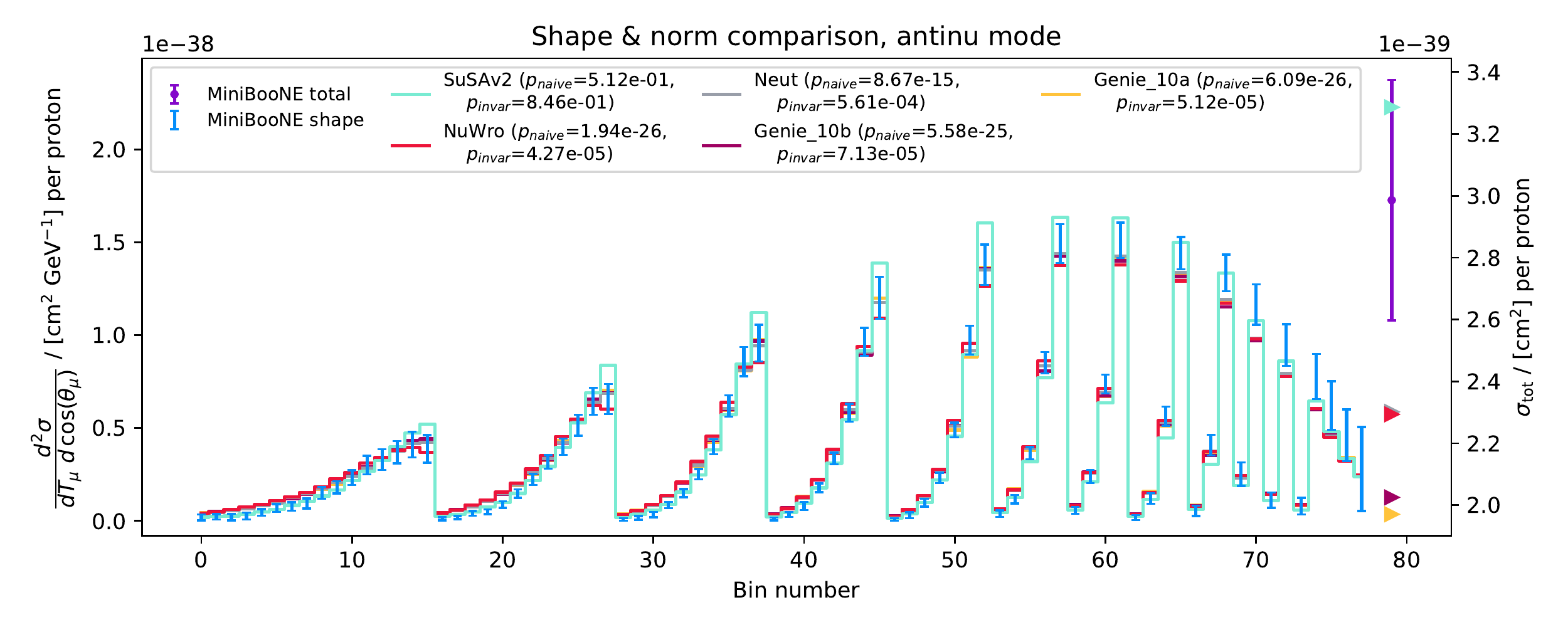}
    \caption{\label{fig:mini-norm}%
    Comparison of shape and normalisation between MiniBooNE data and several model predictions.
    The cosine of the muon angle increases bin-by-bin, the muon's kinetic energy increases block-by-block.
    The error bars show the ``shape error'' of the data.
    The data point on the right shows the common ``normalisation error''.
    The model predictions have been scaled to the same total cross section as the data for the shape comparison,
    i.e. the sum over all bins is identical for all models and the data.
    The points on the right show the actual total cross section of the model predictions.
    The p-values were calculated using all bins and the normalisation.
    See \autoref{tab:mini-p} for shape-only p-values.}
\end{figure*}

To make them comparable to the data, they are split into a shape and a normalisation part as described in~\cite{AguilarArevalo2013}.
The shape part is scaled to the total cross-section of the data, so it can be compared directly to the published data points:
\begin{gather}
    x^{\text{data/MC,norm}} = \sum_i x^\text{data/MC}_i w_i\text{,} \label{eq:norm}\\
    x^{\text{MC,shape}}_i = \frac{x^\text{MC}_i}{x^\text{MC,norm}} x^{\text{data,norm}}\text{,}
\end{gather}
with the 2D bin area $w_i$, which is constant for all bins in this case.
\autoref{fig:mini-norm} shows the comparison of the data and the model predictions.
Despite the lack of covariance information and the implied claim that the uncertainties are independent,
it is clear that the ``shape errors'' must be correlated in some way.
Not only does the cross-section vary too smoothly from bin to bin,
but any variation of only the shape must yield a set of points with a constant sum, meaning the uncertainties of the points cannot be uncorrelated.
Furthermore, it is reasonable to assume that there could be correlations between the normalisation and the shape.
Such correlations would easily arise from scaling uncertainties that affect some bins more than others.

Given the data and the model predictions, it is easy to calculate p-values with the ``naive'' and the ``invariant~3'' test statistics:
\begin{gather}
    p_\text{naive} = 1 - F_{\chi^2_N}\qty(\naive\qty(\bm{\Delta}\,|\,\bm{s}^\text{data}))\text{,} \\
    p_\text{invar} = 1 - F_{\chi^2_1}\qty(\invariant_3\qty(\bm{\Delta} \,|\,\bm{s}^\text{data},\alpha=2/3))\text{,}
\end{gather}
with $\bm{\Delta} = \bm{x}^\text{data} - \bm{x}^\text{MC}$, and the number of data points $N$.
Both the vectors $\bm{x}$ and $N$ explicitly include the added ``normalisation bin'' when comparing both shape and normalisation.

\begin{table*}
    \centering
    \caption{\label{tab:mini-p}%
    P-values from the comparison of models and MiniBooNE data.}
    \begin{tabular}{lll|ccccc}
 & & & Genie 10a & Genie 10b & Neut & NuWro & SuSAv2 \\ \hline
$\nu$ & Shape & naive & 1.89e-03 & 2.29e-03 & 5.24e-10 & 5.14e-06 & 2.19e-24 \\
 & \& norm. & invariant & 2.76e-04 & 2.94e-03 & 1.82e-02 & 1.31e-01 & 2.48e-03 \\
 & Shape  & naive & 3.50e-02 & 2.37e-02 & 5.66e-09 & 2.27e-05 & 3.40e-24 \\
 & only & invariant & 6.89e-02 & 1.40e-01 & 1.81e-02 & 1.30e-01 & 2.46e-03 \\
$\bar\nu$ & Shape & naive & 6.09e-26 & 5.58e-25 & 8.67e-15 & 1.94e-26 & 5.12e-01 \\
 &  \& norm. & invariant & 5.12e-05 & 7.13e-05 & 5.61e-04 & 4.27e-05 & 8.46e-01 \\
 & Shape & naive & 9.92e-24 & 3.71e-23 & 2.82e-14 & 7.32e-26 & 4.96e-01 \\
 & only & invariant & 5.05e-05 & 7.05e-05 & 5.54e-04 & 4.22e-05 & 8.42e-01
    \end{tabular}
\end{table*}

\autoref{tab:mini-p} shows the p-values resulting from the comparisons using  the two test statistics, with and without the normalisation bin.
In most cases, the naive chi-square statistic suggest a much stronger disagreement between the data and the model than the invariant statistic.
This is consistent with the behaviour we have seen in the toy studies.
Apparent strong statistical significance of the naive test statistic, actually corresponds to a much weaker significance in the presence of correlations (see \autoref{fig:naive}),
while the invariant test statistic tends to be conservative (see \autoref{fig:invariant3} and~\ref{fig:realistic}).

\begin{figure*}
    \centering
    \includegraphics[width=\textwidth]{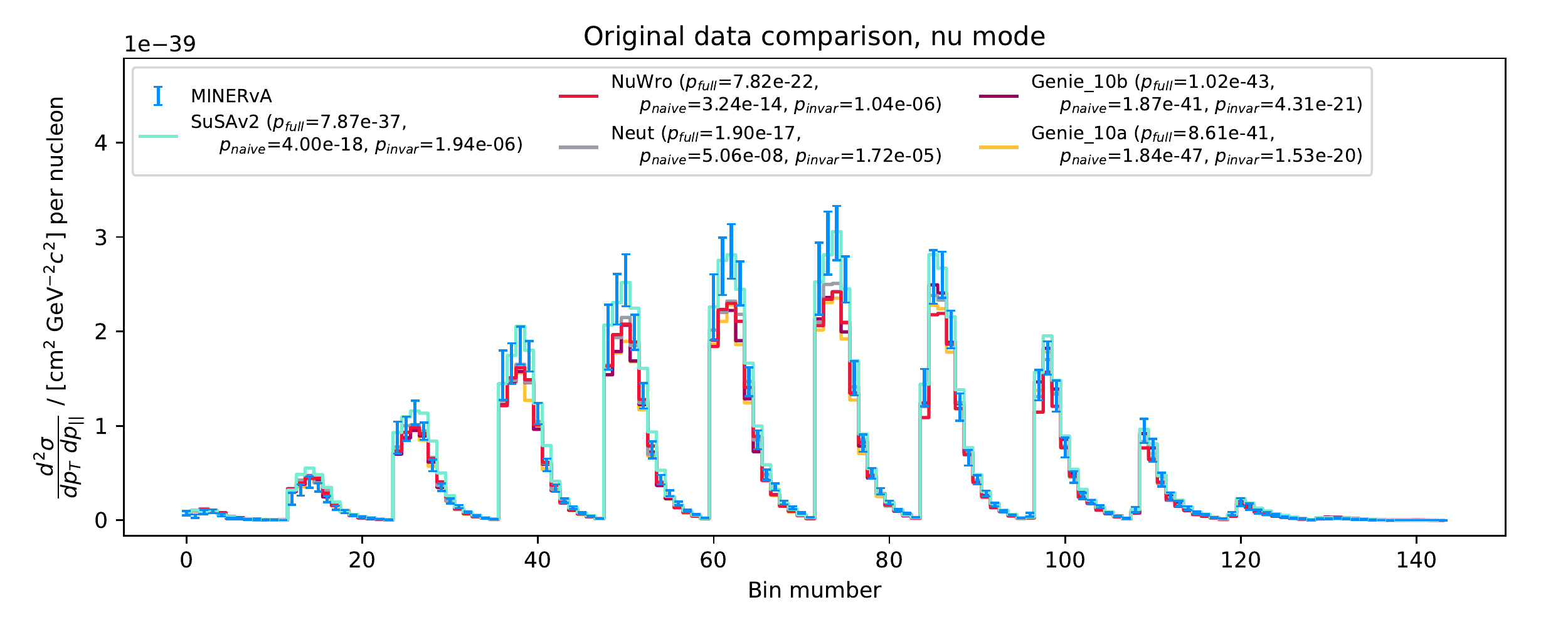}
    \includegraphics[width=\textwidth]{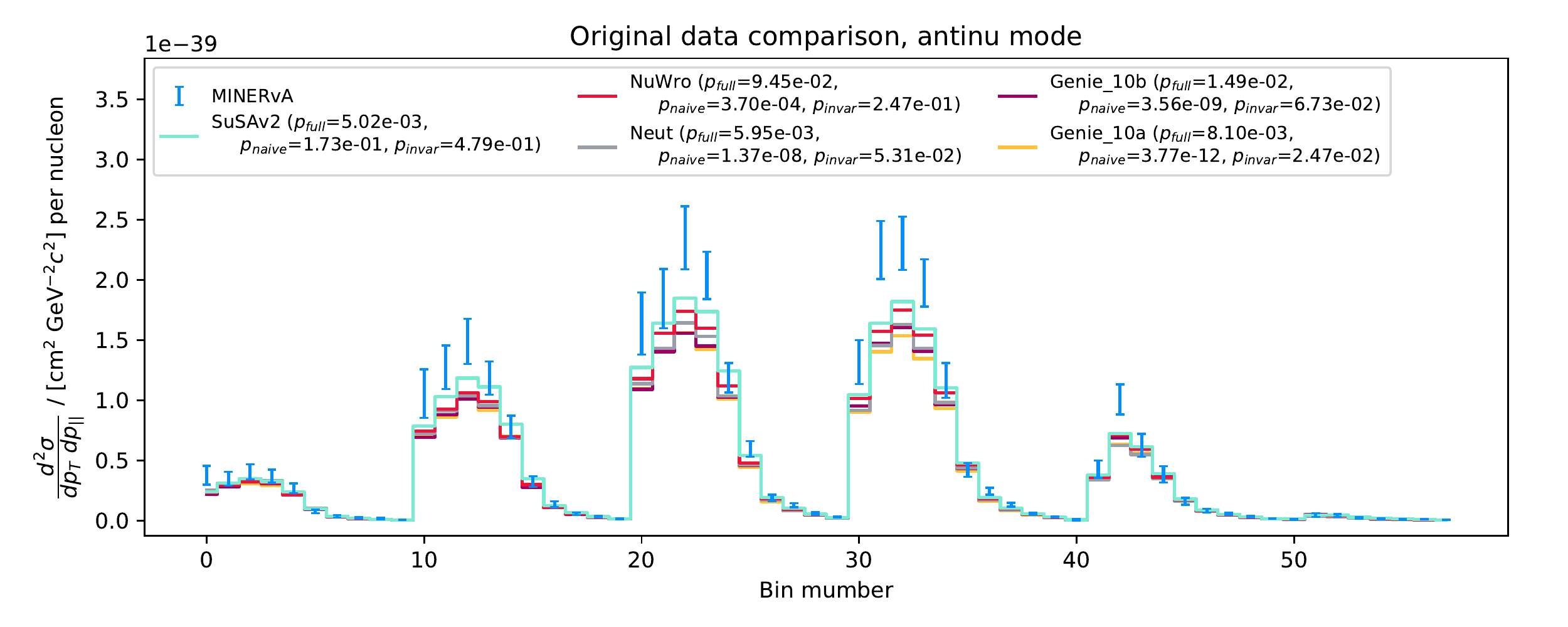}
    \caption{\label{fig:minerva}%
    Comparison between original MINERvA data and several model predictions.
    The muon's parallel momentum $p_{||}$ increases bin-by-bin, its transverse momentum $p_{T}$ block-by-block.
    The error bars show the diagonal elements of the full covariance of the data.}
\end{figure*}

Since the MiniBooNE publications do not provide a full covariance matrix,
it is impossible to judge how the ``diagonal-only'' statistics compare to the ``correct'' results using the information about correlations.
For this we can look at a comparable data set from the MINERvA experiment.
In \cite{Ruterbories2019} and \cite{Patrick2018} they report double-differential, quasielastic-like cross sections in variables of muon momentum.
The number of bins is comparable with the MiniBooNE measurements and they use a similar data unfolding strategy.

Unlike MiniBooNE, MINERvA reports the cross sections with a full covariance matrix, and they do not decompose the uncertainties into a shape and a normalisation part.
This means we can directly compare the p-values we obtain when using the full Mahalanobis distance, with the ones obtained using the test statistics ignoring the off-diagonals of the covariance matrix (see \autoref{fig:minerva}).
None of the considered models describe all of the data particularly well.
The naive test statistic sometimes yields a better and sometimes a worse fit between model and data compared to the full Mahalanobis distance.
The ``invariant~3'' test statistic is consistently conservative.
It is worth noting that while the ``naive'' statistic tends to be closer to the ``correct'' answer for the models with very poor fits, it consistently overestimates the tension between data and model for the better fitting ones, e.g. the NuWro prediction of the antineutrino data.

\begin{figure*}
    \centering
    \includegraphics[width=0.49\textwidth]{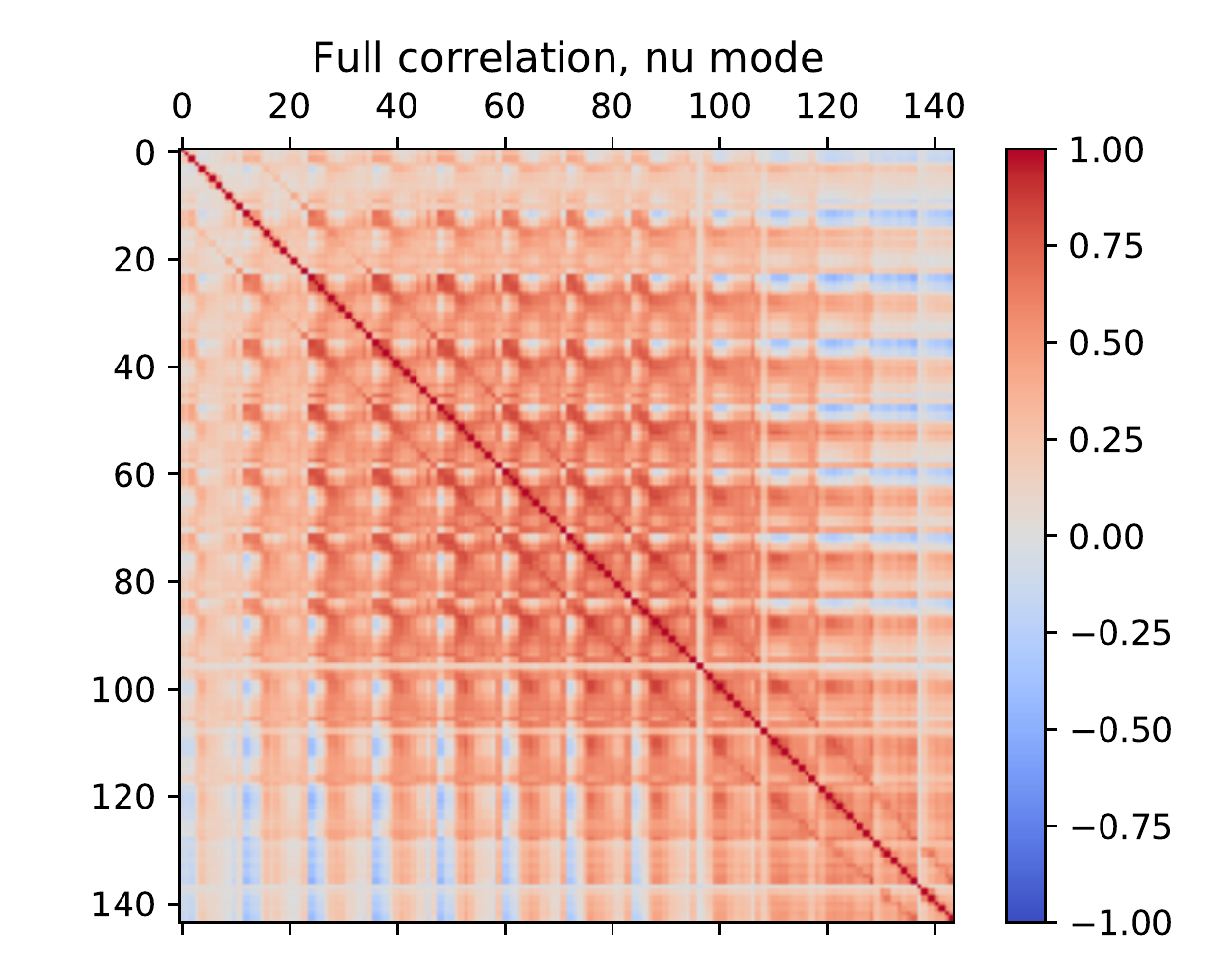}
    \includegraphics[width=0.49\textwidth]{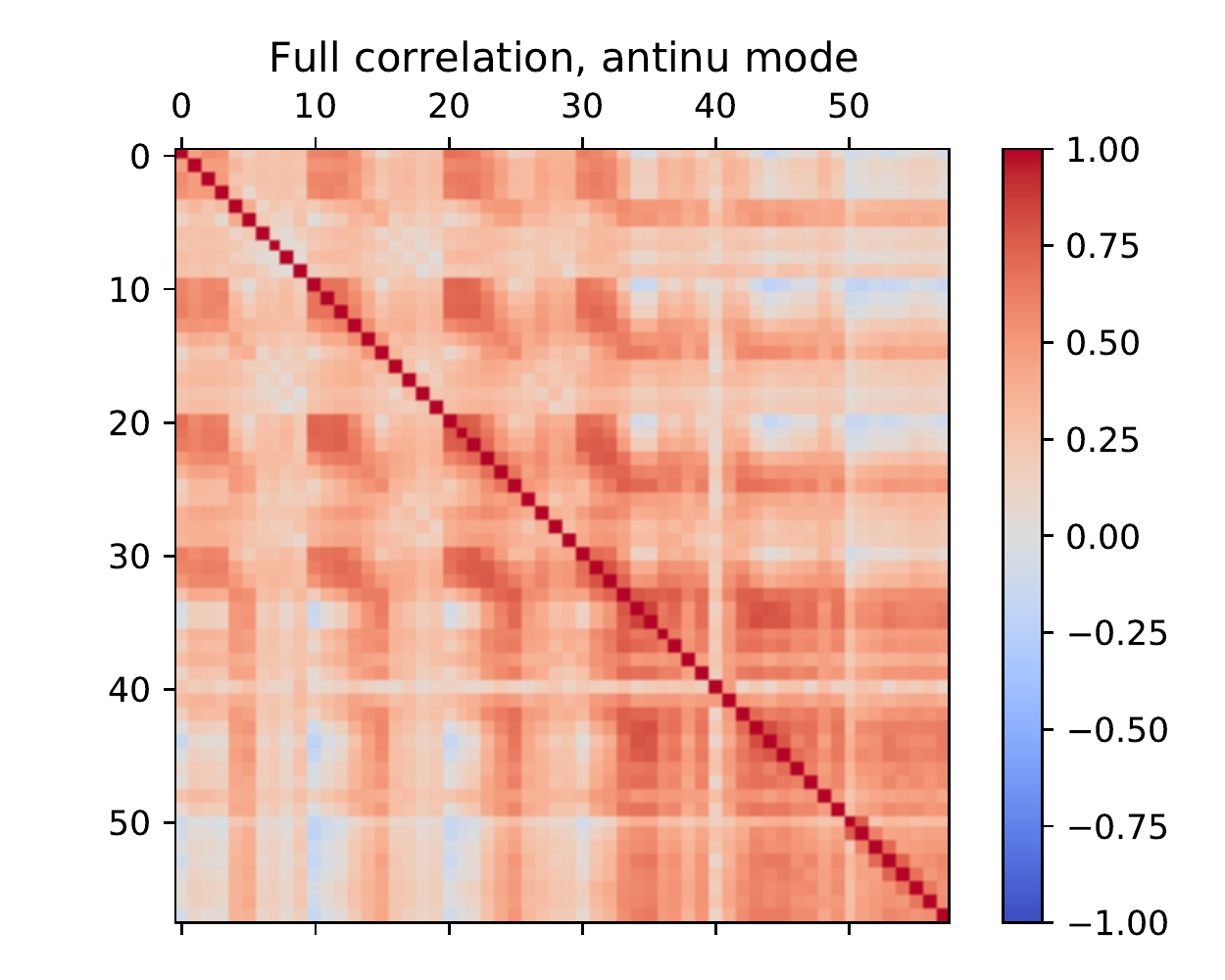}
    \includegraphics[width=0.49\textwidth]{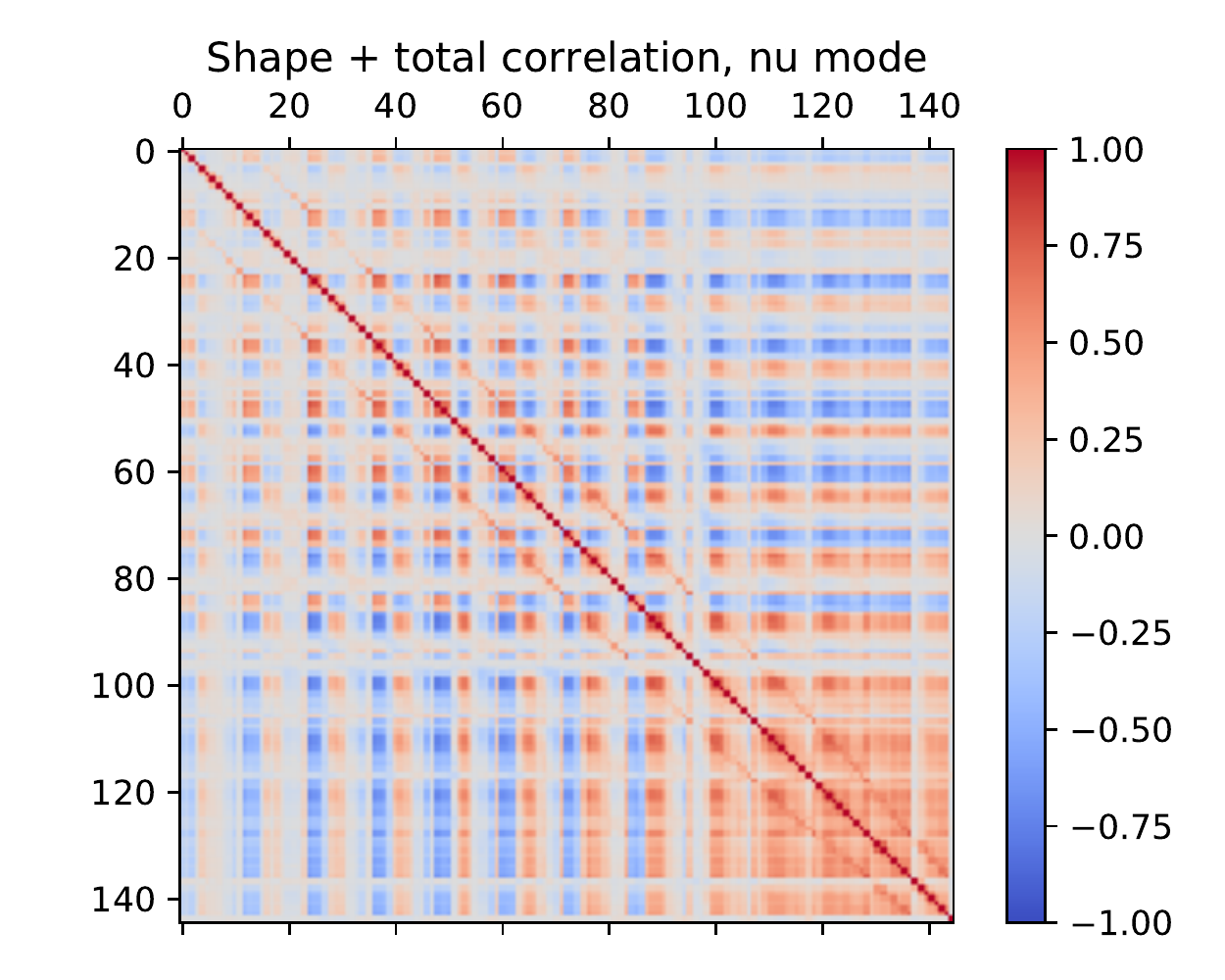}
    \includegraphics[width=0.49\textwidth]{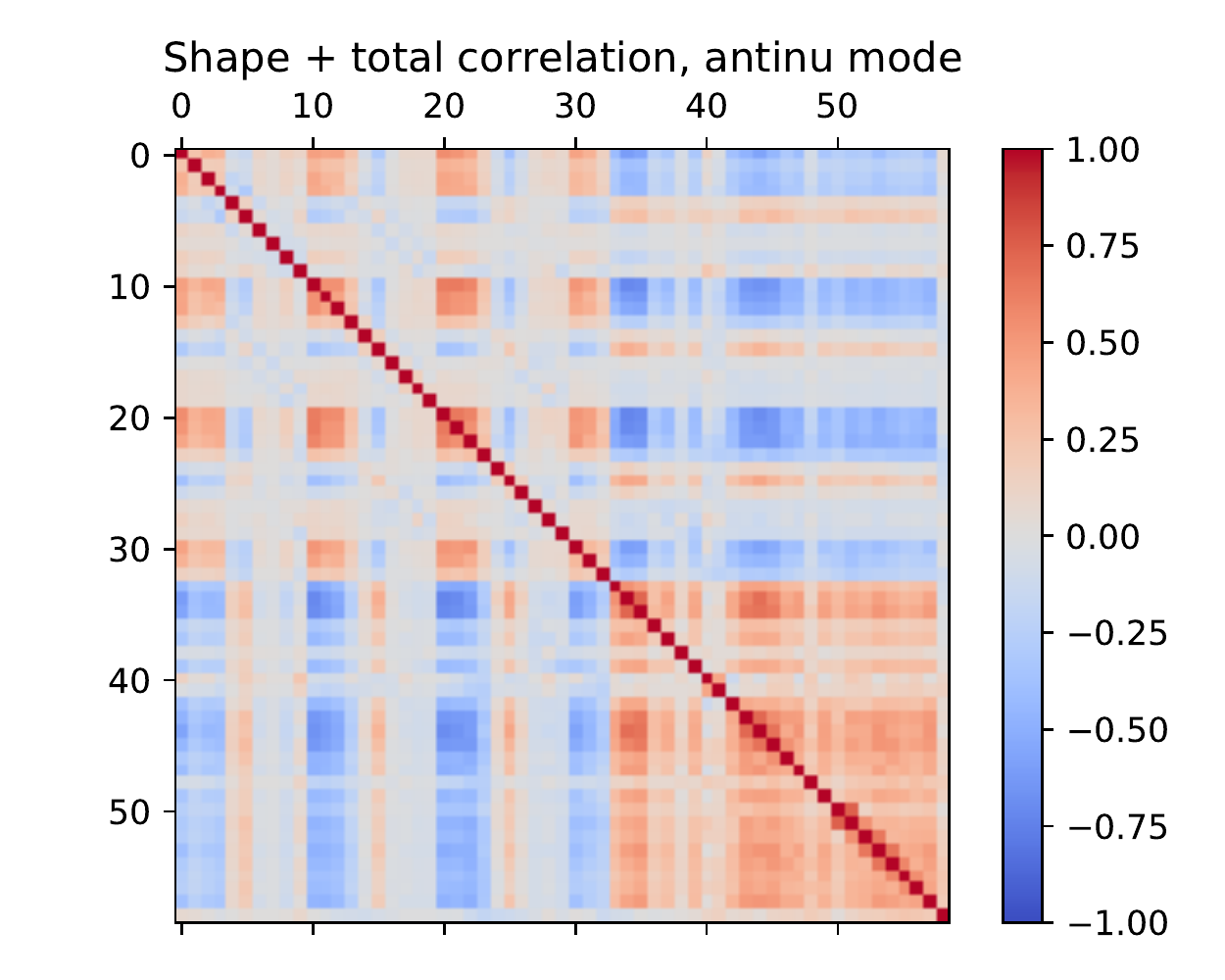}
    \caption{\label{fig:minerva-corr}%
    Correlation matrix of the MINERvA experiment as reported (top) and after decomposition into shape and normalisation bin (bottom) for both neutrino (left) and antineutrino (right) modes.
    The decomposition reduces the strong positive correlations in the data,
    but considerable positive and negative correlations remain.}
\end{figure*}

\begin{figure*}
    \centering
    \includegraphics[width=\textwidth]{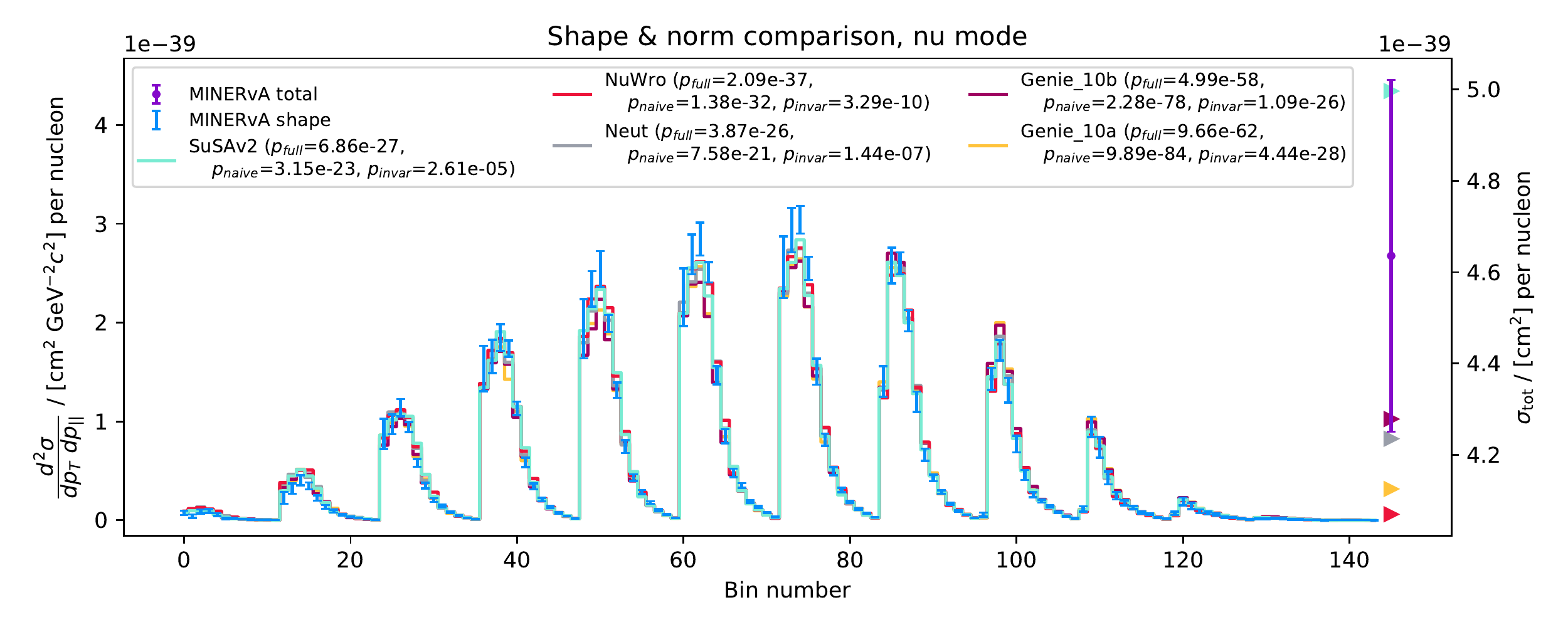}
    \includegraphics[width=\textwidth]{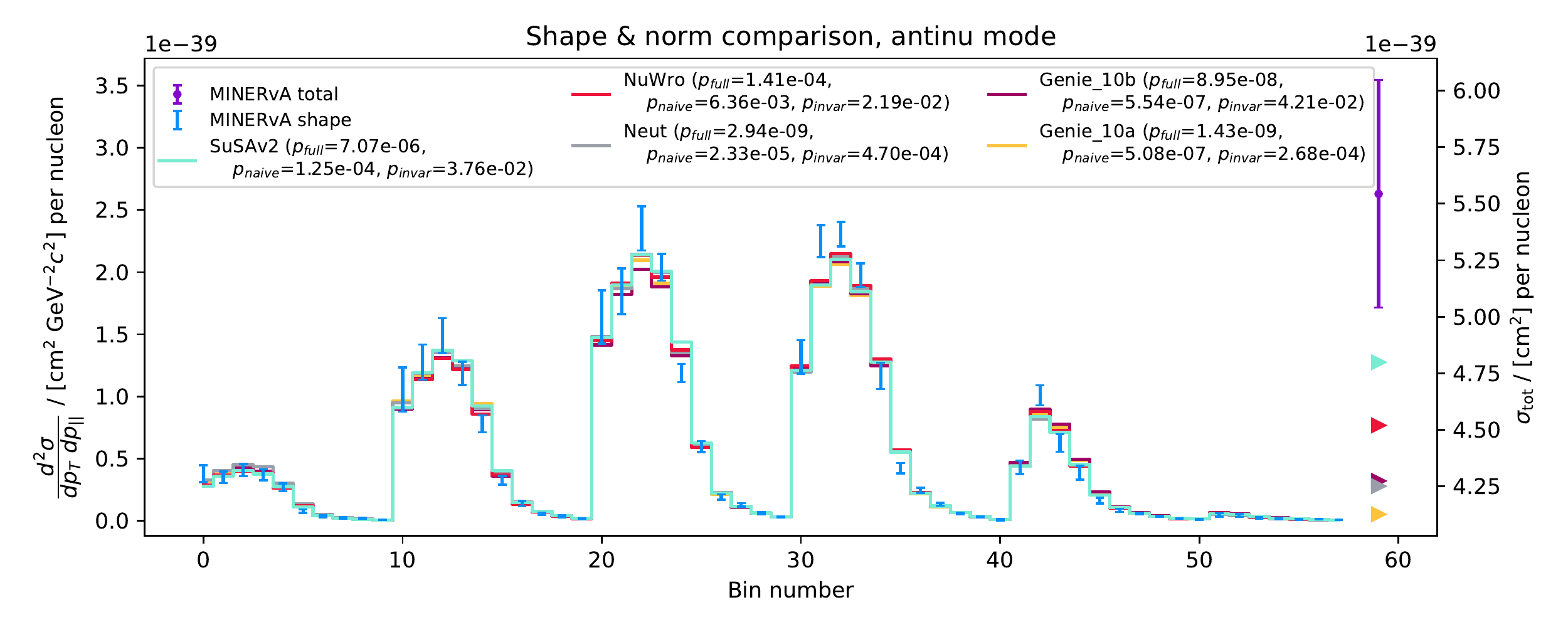}
    \caption{\label{fig:minerva-decomposed}%
    Comparison between decomposed MINERvA data and several model predictions.
    The muon's parallel momentum $p_{||}$ increases bin-by-bin, its transverse momentum $p_{T}$ block-by-block.
    The error bars show the ``shape error'' of the data.
    The data point on the right shows the common ``normalisation error''.
    The model predictions have been scaled to the same total cross section as the data for the shape comparison,
    i.e. the sum over all bins weighted by the 2D~bin widths is identical for all models and the data.
    The points on the right show the actual total cross section of the model predictions.
    The p-values were calculated using all shape bins and the normalisation.
    See \autoref{tab:minerva-p} for shape-only p-values.}
\end{figure*}

We can make the MINERvA data look even more like the MiniBooNE one by decomposing the uncertainty into a shape and a normalisation part following  \autoref{eq:norm}.
Since we have the covariance of the original cross section,
we can also calculate the covariance of the shape and norm:
\begin{equation}
    Q_{ij} = \sum_{k,l} J_{ki} M_{ij} J_{lj} \text{.}
\end{equation}
Here $M$ is the original covariance matrix and $J$ is the Jacobian matrix of the combined shape and norm vector:
\begin{equation}
    J_{ij} =
    \begin{cases}
        \delta_{ij} x_{\text{norm}}^{-1} - x_i w_j x_{\text{norm}}^{-2} & \qif{0 \le i < N} \\
        w_j & \qif{i = N} \text{,}
    \end{cases}
\end{equation}
with the number of original cross-section bins $N$.
The correlation matrices before and after the decomposition are shown in \autoref{fig:minerva-corr}.
As intended, the decomposition reduces the amount of positive correlation in the data,
but considerable positive and negative correlations remain.

\autoref{fig:minerva-decomposed} shows the model comparisons with different test statistics in this decomposed data.
Following MiniBooNE's approach, the shape errors are scaled up by the total cross section in the data for plotting purposes.
Again, the ``invariant~3'' test statistic is consistently conservative compared to the full Mahalanobis distance,
while the ``naive'' test statistic is not.
Note that for the calculation of the Mahalanobis distance, the pseudo-inverse of the covariance matrix was used.
Since shape and norm together have one additional dimension compared to the original cross section, their covariance matrix is not positive definite in general.

\autoref{tab:minerva-p} summarises the p-values from the different model comparisons to the MINERvA data.
The ``full'' p-values using the correlated Mahalanobis distance in the shape and norm case are different from the original case,
despite the fact that they should contain the same information.
This is caused by the different parameterisations of the problem space.
Since the decomposition in shape and norm is not a linear transformation,
the likelihood surfaces described by the covariance matrices cannot be identical.
It is to be expected that this will lead to differing p-values, especially in the low-probability tails.
But even for the comparatively well-fitting models with p-values in the order of a few percent,
the difference is surprisingly strong.
The NuWro prediction of the antineutrino data has a p-value of $\sim 13\%$ in the original paramterisation,
but only $0.03\%$ in the decomposed view.
This would make a huge difference in the interpretation of the data with respect to this model!
Note also, that the invariant test statistic in the decomposed case is much closer to the original full p-values in those cases.

\begin{figure*}
    \centering
    \includegraphics[width=0.49\textwidth]{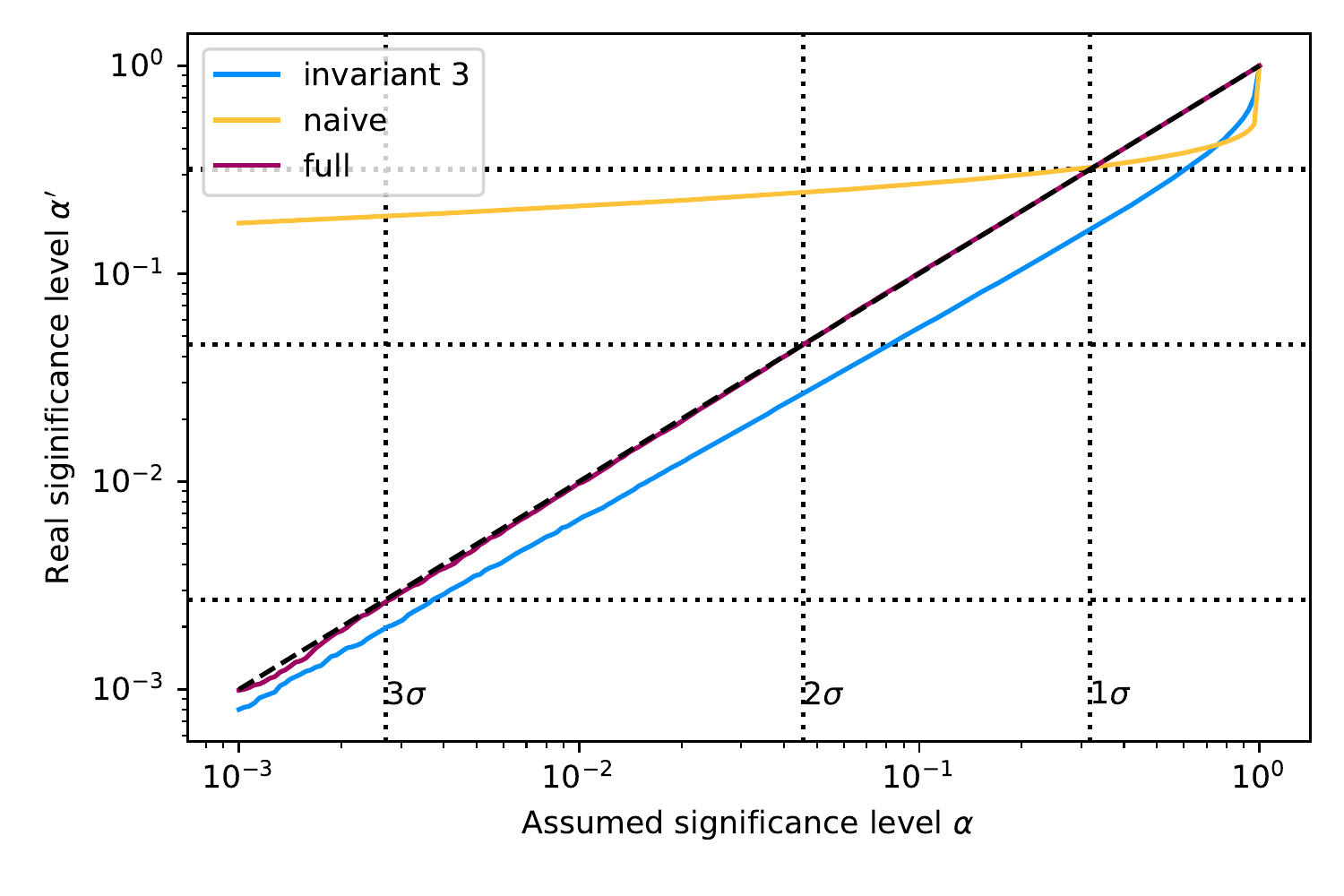}
    \includegraphics[width=0.49\textwidth]{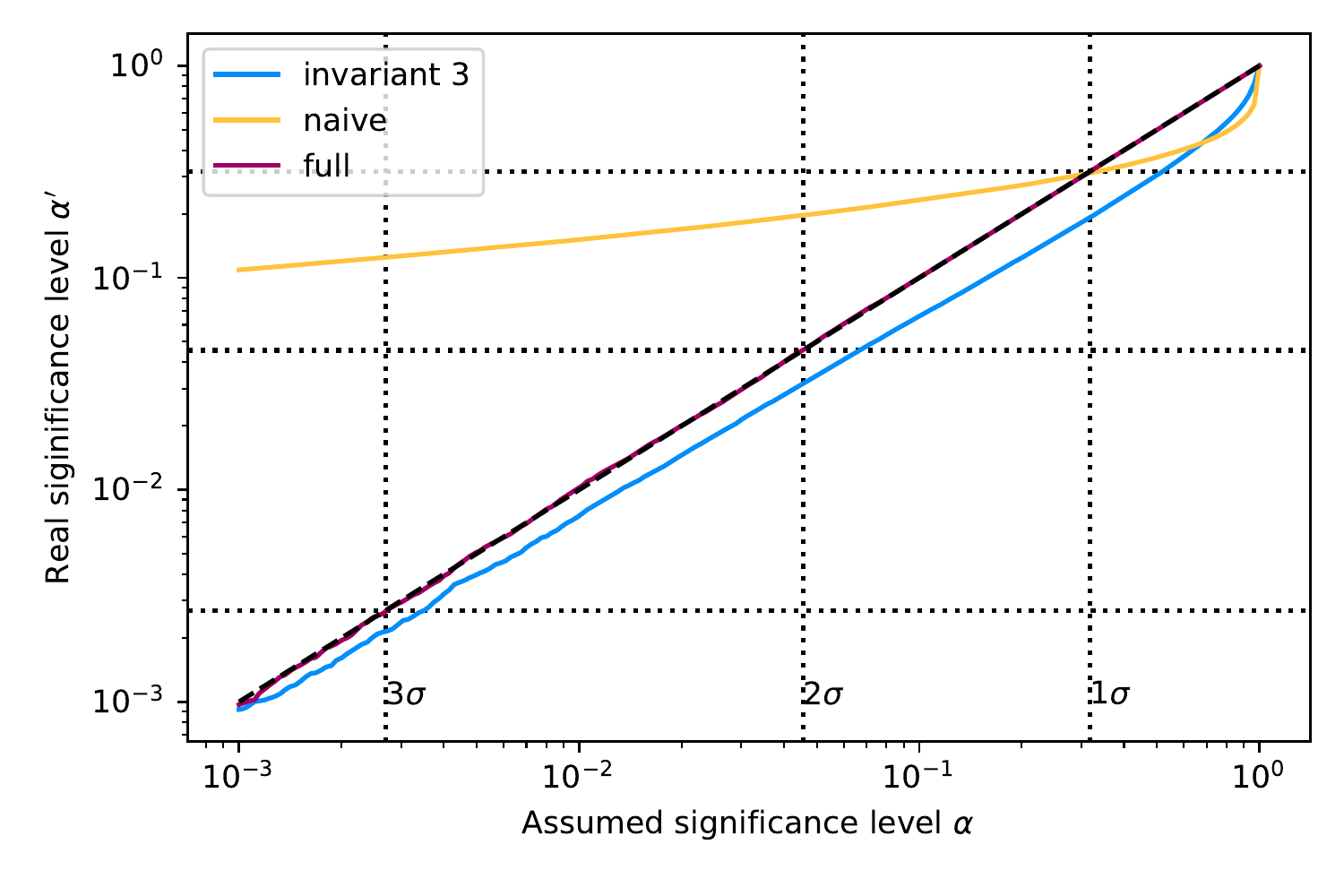}
    \includegraphics[width=0.49\textwidth]{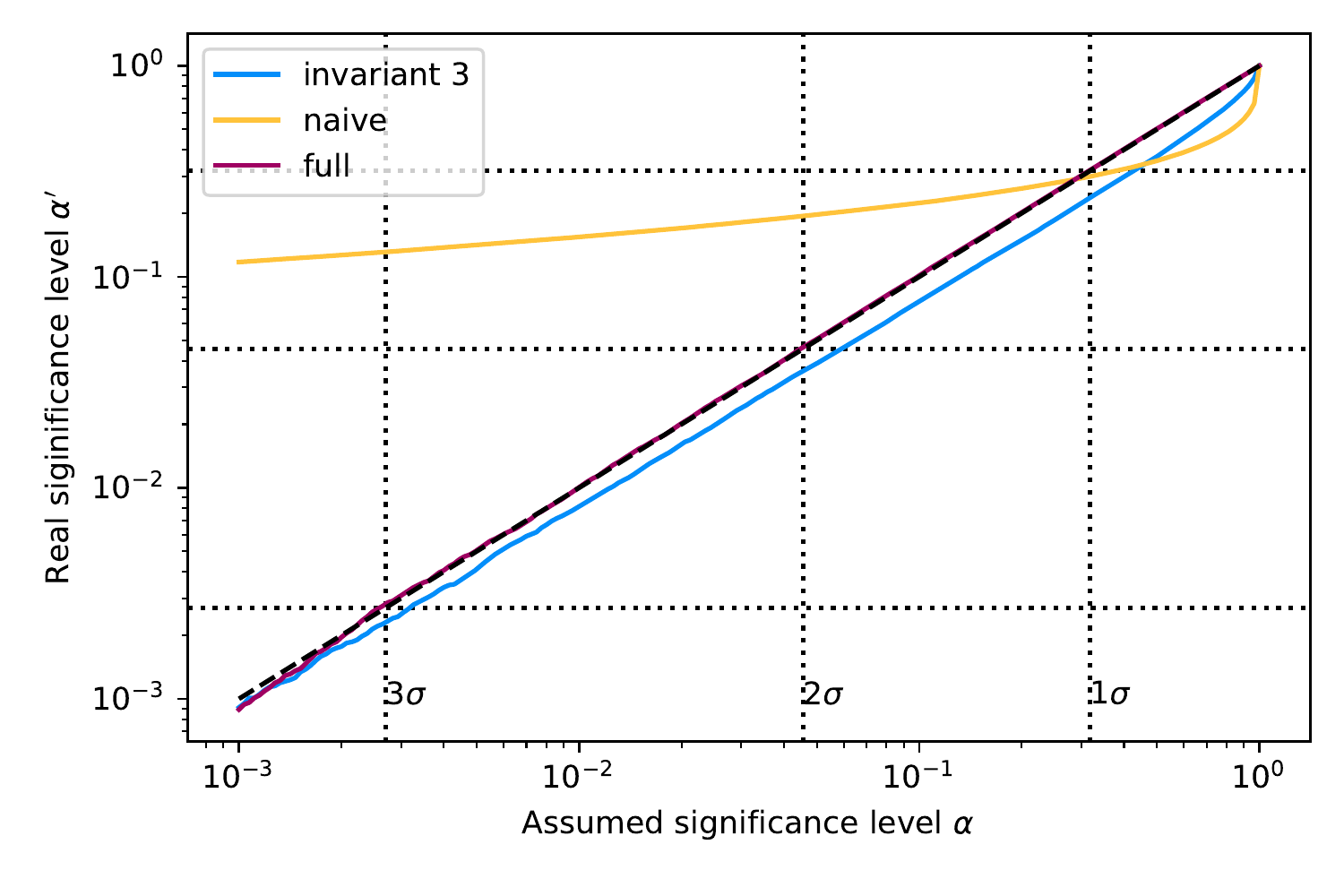}
    \includegraphics[width=0.49\textwidth]{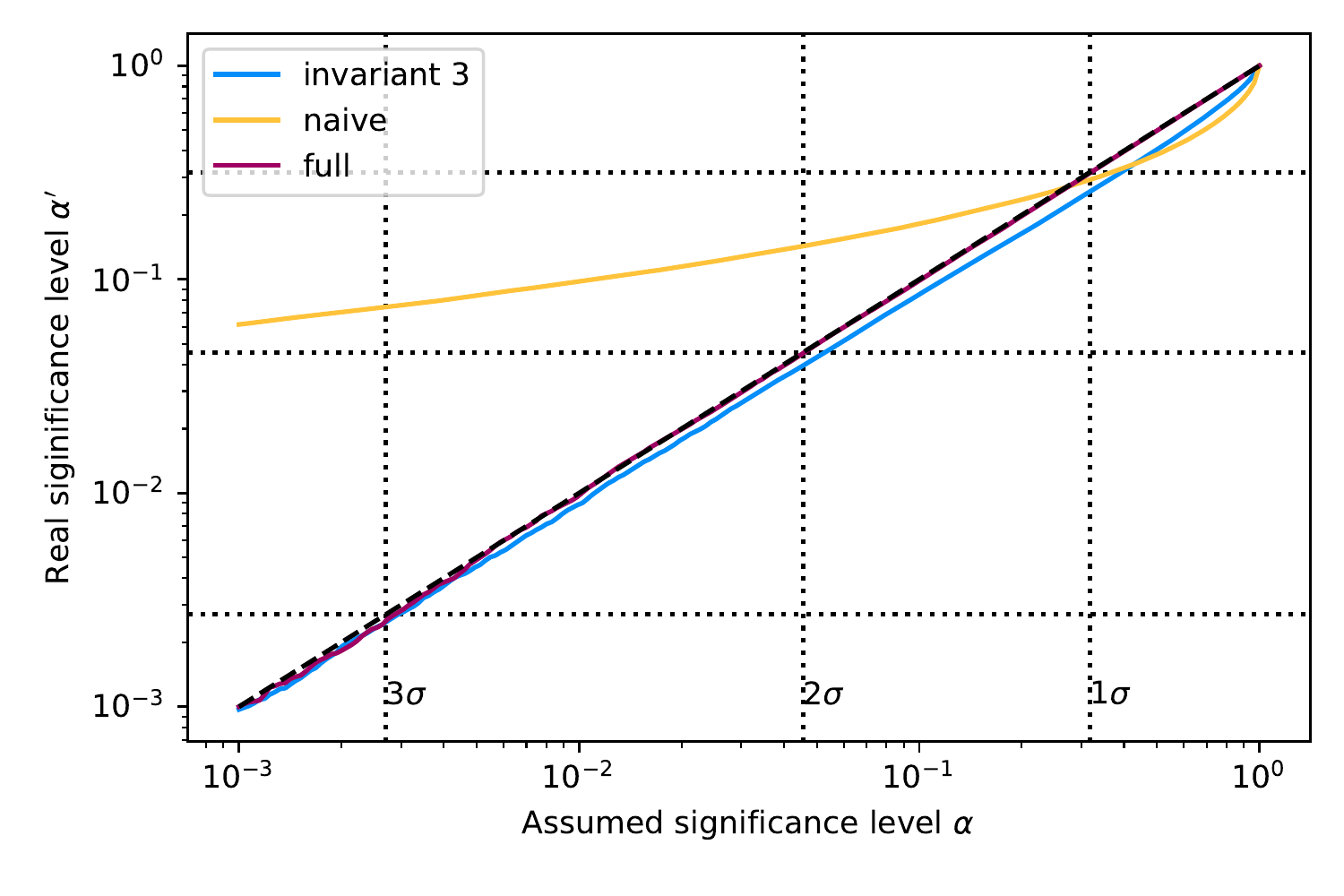}
    \includegraphics[width=0.49\textwidth]{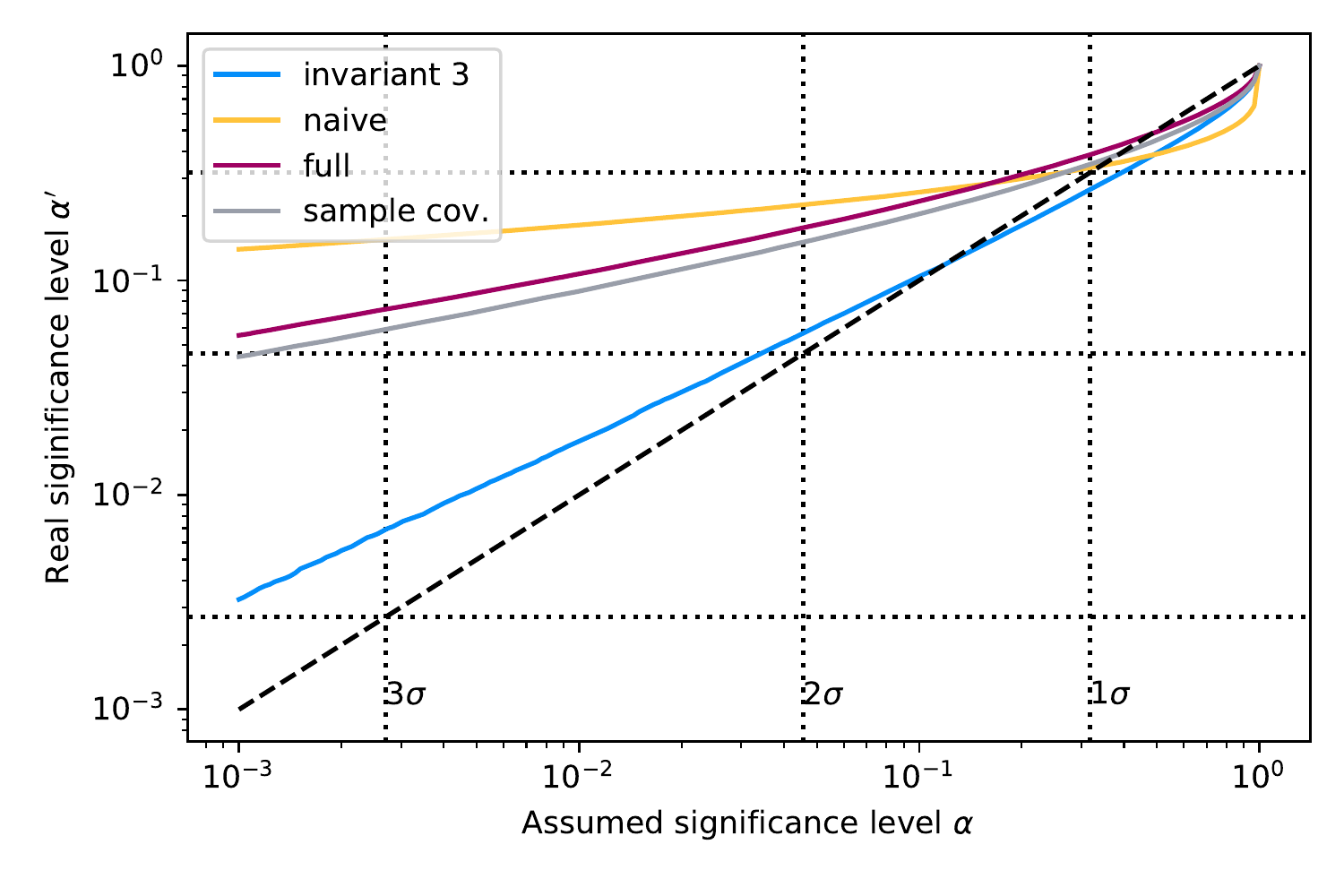}
    \includegraphics[width=0.49\textwidth]{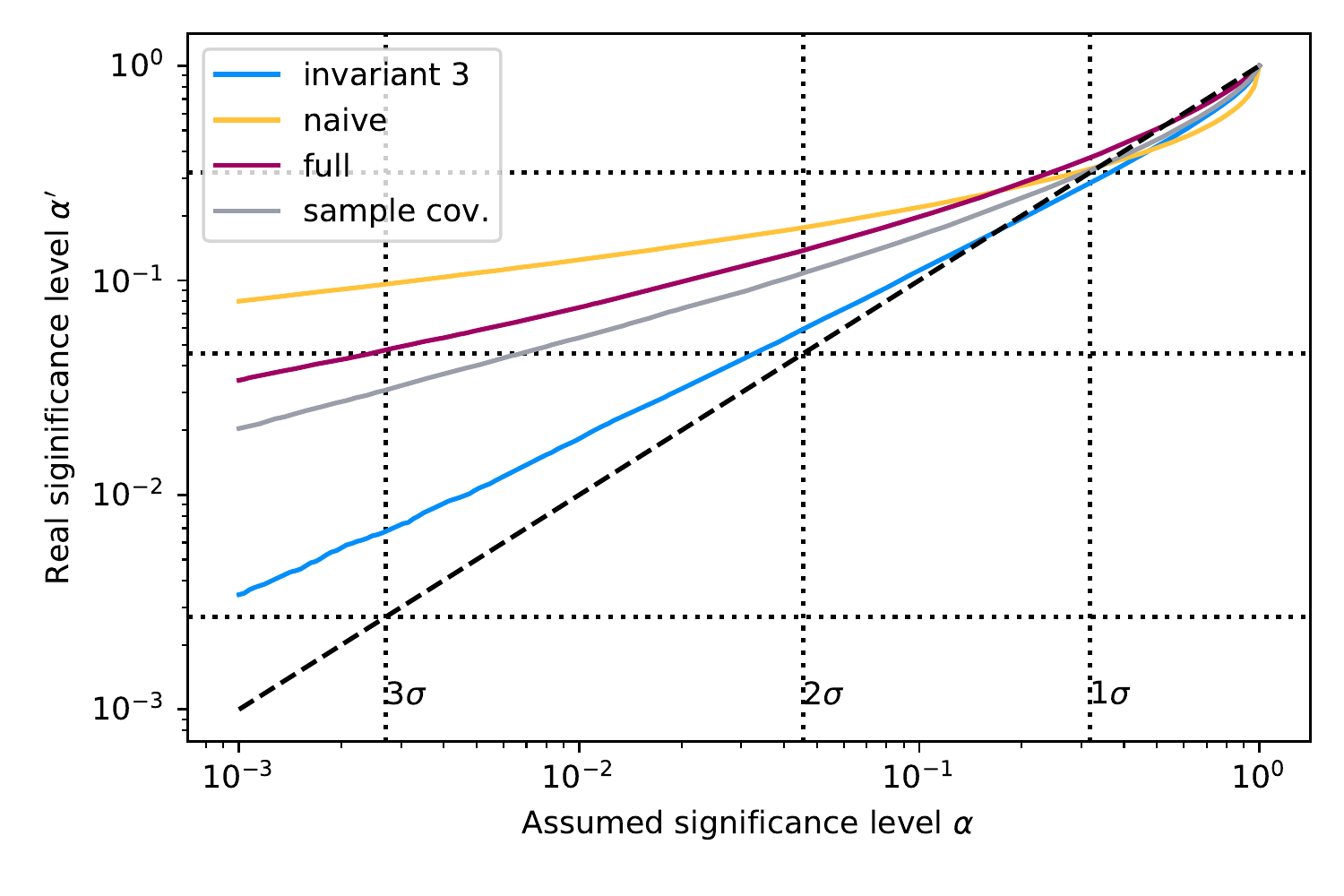}
    \caption{\label{fig:minerva-significance}%
        Assumed vs actual significances with different test statistics in toy data distributed according to the provided MINERvA covariance (top) and the covariance of the decomposition into shape and total cross section (middle) for both the neutrino (left) and antineutrino data (right).
        The bottom plots show the calculated and true significance levels when generating toy data according to the original covariance and applying the decomposition afterwards.
        The nonlinear transformation of the decomposition causes a significant distortion in the distribution of the Mahalanobis distance, even when using the full covariance matrix.
        This improves only slightly when using the sample covariance instead of the linear error propagation.
        The invariant test statistic seems to be less sensitive to these distortions than the full Mahalanobis distance.
    }
\end{figure*}

Since we have the covariance matrices, we can create toy data that is distributed according to them and check for the coverage properties of the different test statistics, just like we did in the previous sections.
The results of these studies is shown in \autoref{fig:minerva-significance}.
The top plot shows the performance of the full, naive, and invariant statistics when applied to the data distributed according to MINERvA's original covariance matrix.
The middle plots show the same for data distributed according to the decomposed covariance.
As in the previous studies, we see that the invariant test statistic shows much more accurate coverage properties than the naive one.
The full Mahalanobis distance performs best, as expected.

In the bottom plots however, we see the performance of the three statistics when generating data sets according to the original covariance, and then applying the decomposition.
Because of the non-linear transformation of the decomposition,
the covariance of the shape \& norm parameterisation does no longer reflect the actual distribution of the data.
This deteriorates the coverage properties of the full Mahalanobis distance test statistic to the point where it performs as bad as, or worse than the naive test statistic.
Interestingly, the invariant test statistic is not as affected by the non-linear transformation as the the other two.
It is no longer conservative, but it shows better coverage properties than even the full statistic.
This is probably due to the fact that the full statistic is trying to make full use of the assumed shape of the data distribution, i.e. the correlation between the data points,
which is distorted due to a non-linear transformation.
The invariant test statistic on the other hand is designed to work well with a wide range of correlations, in the data.
Note that the performance of the full Mahalanobis distance improves slightly when calculating the covariance in the decomposed space directly from the sample, rather than doing linear error propagation.
It still performs rather poorly, because of the non-gaussian shape of the underlying data, though.
This underlines the importance of checking that reported uncertainties are actually sufficiently normal distributed in the chosen parameterisation,
when reporting them as a covariance matrix.

\begin{table*}
    \centering
    \caption{\label{tab:minerva-p}%
    P-values from the comparison of models and MINERvA data.}
    \begin{tabular}{lll|ccccc}
 & & & Genie 10a & Genie 10b & Neut & NuWro & SuSAv2 \\ \hline
$\nu$ & Original & full & 8.61e-41 & 1.02e-43 & 1.90e-17 & 7.82e-22 & 7.87e-37 \\
 &  & naive & 1.84e-47 & 1.87e-41 & 5.06e-08 & 3.24e-14 & 4.00e-18 \\
 &  & invariant & 1.53e-20 & 4.31e-21 & 1.72e-05 & 1.04e-06 & 1.94e-06 \\
 & Shape & full & 9.66e-62 & 4.99e-58 & 3.87e-26 & 2.09e-37 & 6.86e-27 \\
 & \& norm. & naive & 9.89e-84 & 2.28e-78 & 7.58e-21 & 1.38e-32 & 3.15e-23 \\
 &  & invariant & 4.44e-28 & 1.09e-26 & 1.44e-07 & 3.29e-10 & 2.61e-05 \\
 & Shape & full & 9.37e-62 & 4.32e-58 & 3.11e-26 & 1.96e-37 & 7.36e-27 \\
 & only  & naive & 8.78e-84 & 1.43e-78 & 6.60e-21 & 1.63e-32 & 2.54e-23 \\
 &  & invariant & 4.41e-28 & 1.08e-26 & 1.43e-07 & 3.27e-10 & 2.60e-05 \\
$\bar\nu$ & Original & full & 8.10e-03 & 1.49e-02 & 5.95e-03 & 9.45e-02 & 5.02e-03 \\
 &  & naive & 3.77e-12 & 3.56e-09 & 1.37e-08 & 3.70e-04 & 1.73e-01 \\
 &  & invariant & 2.47e-02 & 6.73e-02 & 5.31e-02 & 2.47e-01 & 4.79e-01 \\
 & Shape & full & 1.43e-09 & 8.95e-08 & 2.94e-09 & 1.41e-04 & 7.07e-06 \\
 & \& norm.  & naive & 5.08e-07 & 5.54e-07 & 2.33e-05 & 6.36e-03 & 1.25e-04 \\
 &  & invariant & 2.68e-04 & 4.21e-02 & 4.70e-04 & 2.19e-02 & 3.76e-02 \\
 & Shape & full & 6.04e-08 & 2.71e-06 & 1.28e-08 & 2.78e-04 & 9.13e-06 \\
 & only  & naive & 3.10e-06 & 2.19e-06 & 8.74e-05 & 1.12e-02 & 1.54e-04 \\
 &  & invariant & 2.64e-04 & 4.14e-02 & 4.62e-04 & 2.15e-02 & 3.69e-02
    \end{tabular}
\end{table*}

\section{Conclusions}

We have shown that the use of the ``naive'' uncorrelated Mahalanobis distance in the presence of unknown correlations leads to wrong coverage properties.
The presented alternative test statistics perform more robust under varying degrees of correlation in the data.

The ``fitted'' test statistic is motivated by treating the correlations as nuisance parameters.
Its distribution in the absence of correlations is known (the ``Bee-square distribution'') and it is conservative in their presence.
It could be used when it is important to have an ``actual'' Mahalanobis distance\footnote{The usefulness of this is probably limited though, as it will not be chi-square distributed in general.} for the combination with other data sets,
and the overestimation of the error in the presence of correlations is acceptable.
It also has the advantage of being incredibly easy to calculate,
as it is just the maximum squared z-score among the variables.
Depending on the actual correlations in the data, its performance is actually comparable to the ``invariant'' test statistics.

The ``invariant~3'' test statistic performs the best across varying levels of correlations.
Its level of conservativeness can be tuned with the shape parameter $\alpha$.
A value of 0.5 seems to be a safe choice, while the best performance in the toy data sets in the presented studies was achieved at $\alpha = 2/3$.
As it goes to 0, the test statistic becomes equivalent to the ``invariant~2'' statistic,
which can be seen as the most conservative version of ``invariant~3''.
The ``invariant~3'' statistic could be used when it is important to not over-estimate the uncertainties by too much.
If necessary, the relative weight of the statistic can be tuned by transforming it to a chi-square distribution with $N > 1$ degrees of freedom.
The choice of $N$ is somewhat arbitrary though.

The application to real data from MiniBooNE and MINERvA shows that the ``invariant~3'' test statistic performs as expected.
It is consistently conservative, while the ``naive'' uncorrelated Mahalanobis distance overestimates the strength of the discrepancy between data and model in multiple instances.
When the MINERvA data is decomposed into a shape and a normalisation part,
the invariant statistic even has better coverage properties than a full Mahalanobis distance that uses a linearly propagated covariance matrix.
This is most likely due to the non-linear nature of the decomposition into shape and norm,
which distorts the shape of the distribution.
If it was multivariate Gaussian originally, the covariance in the shape and norm space can only ever be an approximation.
Since the invariant statistic does not use information about the correlations between the data points,
it is not affect as much by this as the fully correlated Mahalanobis distance.

\section*{Thanks}

I would like to thank Callum Wilkinson and Stephen Dolan for generating the NUISANCE files that were used in the MiniBooNE and MINERvA studies, as well as providing a space for discussions, feedback, and venting.
Thanks also goes to Louis Lyons for providing feedback and clarifying discussions about the concepts discussed in this paper.
This work was supported by a grant from the Science and Technology Facilities Council.

\bibliography{biblio}

\appendix

\floatstyle{plaintop}
\newfloat{listing}{p}{lop}
\floatname{listing}{Listing}

\begin{listing*}
\caption{\label{lst:bee}%
Python implementation of the Bee-square distribution.}
\begin{lstlisting}[language=Python,frame=L]
import numpy as np
from scipy.stats import rv_continuous
from scipy.special import erf, erfinv

class Bee(rv_continuous):
    def _cdf(self, x, df):
        return erf(x/np.sqrt(2))**df
    
    def _pdf(self, x, df):
        ret = df*(erf(x/np.sqrt(2)))**(df-1)
        return ret * np.sqrt(2/np.pi)*np.exp(-x**2/2)
    
    def ppf(self, x, df):
        return erfinv((x)**(1/df)) * np.sqrt(2)

# Instance of the distribution, support starts at 0
bee = Bee(a=0)

class Bee2(rv_continuous):
    def _cdf(self, x, df):
        b = np.sqrt(x)
        ret = bee.cdf(b, df)
        return ret

    def _pdf(self, x, df):
        ret = df*(erf(np.sqrt(x/2)))**(df-1)
        return ret / np.sqrt(2*np.pi*x) * np.exp(-x/2)

    def _ppf(self, x, df):
        b = bee.ppf(x, df)
        return b**2

# Instance of the distribution, support starts at 0
bee2 = Bee2(a=0)
\end{lstlisting}
\end{listing*}

\begin{listing*}
\caption{\label{lst:invariant}%
Python implementation of the ``invariant~3'' test statistic.
Data must be provided in z-scores, i.e. normalised by the variance.
Uses caching for improved performance and works with survival functions instead of CDFs for improved numerical accuracy.}
\begin{lstlisting}[language=Python,frame=L]
@np.vectorize
@lru_cache(10000)
def _yfrommax(b, df=2, alpha=0.5):
    """(1 - diagonal coordinate) from (1 - max) of accepted region"""
    # A = (1-b)**df - ((y-b)**df)/((1-alpha+alpha*y)**(df-1)) = 1 - y
    beta = 1 - alpha
    q = (1.0 - b) ** df - 1
    dfm = df - 1

    def f(y):
        return q - ((y - b) ** df) / ((beta + alpha * y) ** (dfm)) + y

    if b <= 0:
        return 0.0
    if b >= 1:
        return 1.0
    else:
        return root_scalar(f, x0=b, x1=b * 1.001).root

def yfrommax(b, df=2, alpha=0.5):
    """Buffer and interpolate values to speed things up."""
    step = 0.0001
    b_ = np.floor(b / step, dtype=float) * step
    b__ = b_ + step
    delta = (b - b_) / step
    x_ = _yfrommax(b_, df=df, alpha=alpha)
    x__ = _yfrommax(b__, df=df, alpha=alpha)
    return x_ + (x__ - x_) * delta

def invariant3(x, alpha=0.5, fast=False):
    """Return test statistic given vector of normalised values."""
    if fast:
        sf = 1 - chi2.cdf(x ** 2, df=1)  # Faster, but less accurate
    else:
        sf = chi2.sf(x ** 2, df=1)

    # Get possible diagonal coordinate from maximum CDF value (= minimum SF)
    a = np.min(sf, axis=-1)
    b = np.max(sf, axis=-1)
    yfm = yfrommax(a, df=x.shape[-1], alpha=alpha)

    # Get possible diagonal coordinate from centre surface
    yfc = (alpha * (a - b) + b) / (1.0 + alpha * (a - b))

    y = np.minimum(yfc, yfm)
    y = np.maximum(y, 0)  # Cap in case of rounding or root finding errors
    return chi2.isf(y, df=1)
\end{lstlisting}
\end{listing*}

\section{Proof of conservativeness of fitted test statistic in two dimensions}
\label{sec:fitted-proof}

The conservativeness of the fitted test statistic can be proven by showing that the CDF of any given maximum absolute value $b$ among the $N$ standard normal distributed variables is minimal, when the variables are uncorrelated.
In two dimensions this can be done by proving that
\begin{equation}\label{eq:2DCDF}
    \qty|S|^{-\frac{1}{2}}\int_{-b}^{b}\int_{-b}^{b}\exp(-\frac{1}{2}\bm{z}^T S^{-1} \bm{z}) \dd{z_1}\dd{z_2}\text{,}
\end{equation}
with $S=\smqty*(1&a \\ a&1)$ and $a \in (-1,1)$, is minimal for $a=0$.

Differentiating term~\ref{eq:2DCDF} with respect to $a$ yields:
\begin{equation}
    \frac{2 \left(1 - e^{\frac{2 a b^{2}}{a^{2} - 1}}\right) e^{\frac{b^{2} \left(1 - a\right)}{a^{2} - 1}}}{\sqrt{1 - a^{2}}}\text{,}
\end{equation}
as can be shown e.g. with computer algebra systems.
This term is positive when $a > 0$, $0$ when $a=0$, and negative when $a<0$.
Thus term~\ref{eq:2DCDF} is minimal at $a=0$.
QED.

This calculation will remain true when integrating over additional statistically independent variables in term \ref{eq:2DCDF}.
Those will add constant factors to the derivative,
but not change the general dependence on $a$.
Starting from a covariance with no correlations,
the CDF is thus globally minimal vs. variations of any single of the off-diagonal elements.

\end{document}